\documentclass[twocolumn]{aastex61}
\usepackage{amsmath,amssymb,amsthm}
\usepackage{wasysym}
\usepackage{bm}
\usepackage{parskip}

\usepackage[normalem]{ulem}

\usepackage{graphicx}                 
\usepackage{xcolor}
\usepackage{tikz}
\usetikzlibrary{fpu,calc,intersections,through,backgrounds}
\usepackage{wrapfig}
\usepackage{MnSymbol}

\received{November 5, 2018}
\revised{December 18, 2018}
\accepted{\today}
\submitjournal{ApJ}

\begin{document}

\title{Systematic structure and sinks in the YORP effect}

\correspondingauthor{Oleksiy Golubov}
\email{oleksiy.golubov@karazin.ua}

\author[0000-0002-5720-2282]{Oleksiy~Golubov}
\affil{Department of Aerospace Engineering Sciences, University of Colorado at Boulder, 429 UCB, Boulder, CO, 80309, USA}
\affiliation{School of Physics and Technology, V. N. Karazin Kharkiv National University, 4 Svobody Sq., Kharkiv, 61022, Ukraine}
\affiliation{Institute of Astronomy of V. N. Karazin Kharkiv National University, 35 Sumska Str., Kharkiv, 61022, Ukraine}

\author[0000-0003-0558-3842]{Daniel J. Scheeres}
\affil{Department of Aerospace Engineering Sciences, University of Colorado at Boulder, 429 UCB, Boulder, CO, 80309, USA}

\begin{abstract}

There is a correlation between the components of the YORP effect of most asteroids, which drives the obliquity and spin rate of the affected bodies in a consistent pattern. 
This allows for a clear and unambiguous picture for how the spin rates and poles of asteroids affected by YORP will evolve and simplifies the overall picture for how populations will migrate on average.

The YORP effect can also lead to a previously unexplored equilibrium state for affected bodies. This equilibrium state is a function of the usual ``normal YORP'' effect (which arises due to the global shape asymmetry of the asteroid) and the ``tangential YORP'' (which arises due the transport of thermal energy through rocky surface features). Estimates from current shape models show that 10-20\% of asteroids have the proper condition to be captured in this equilibrium state, indicating that the occurrence of this state may be significant. The existence of this attractor for the asteroid population means that objects affected by YORP may leave their usual YORP cycles and maintain a constant spin rate over long time periods -- this has significant implications for our interpretation of asteroid spin rate evolution and related theories for their physical evolution.

\end{abstract}


\section{Introduction}
The YORP effect, which dominates the physical evolution of asteroids less than $\sim$10 km in size \citep{rubincam00,bottke06,vokrouhlicky15}, has been characterized as leading to a randomization of asteroid spin rates and spin poles \citep{pravec08}, however we show that the effect has a previously unrealized systematic structure. The YORP effect is defined as torques due to the reemission of incident solar photons from an aspherical, spinning body. It has been implicated as the main cause for the spin acceleration and deceleration of small asteroids, and is known to drive the spin poles of these bodies towards obliquities of 0, 90 or 180 degrees. The effect has been considered to be secular, meaning that there are no physical ``stops'' for the spin rate evolution, so that bodies will tend to spin faster until they disrupt or deform -- potentially changing the sign of the YORP effect, or spin slower until they tumble and become subject to irregular perturbations by gravitational and light pressure torques before they eventually spin up again \citep{pravec08,rossi09}. It has also been considered to be random, in that the combination of spin rate evolution and obliquity evolution are uncorrelated. The migration of an asteroid's spin state through these YORP cycles has been viewed as one of the constants of a small asteroid's life, repeating ad-infinitum and influenced by random shifts in the body's shape \citep{statler09}. 

In this article, we show that a more careful evaluation of the fundamental physics of the YORP effect, including heat transfer effects on the surface, cause a strong correlation in the obliquity and spin rate dynamics which will shape their rotational evolution, and that there are ``escape hatches'' from the YORP cycle in terms of stable equilibrium states that an asteroid's spin state can migrate into. Estimates based on our analytical theory show that between 10-20\% of current asteroid shape models allow for such equilibrium states, and that over time (and repeated YORP cycles) it should be possible for asteroids to naturally land in such a state -- which would then remove them from their Sisyphean fate. These new realizations require theories of the physical evolution of small bodies to be revised, and will strongly influence many of the secondary effects of YORP that are used as a lens to interpret the current small body population. The found equilibria between the normal YORP and the tangential YORP are complementary to other previously discussed types of equilibria created by the normal YORP in the presence of tumbling \citep{breiter15} or thermal lag \citep{scheeres08}, as well as in binary asteroid systems \citep{golubov16equilibrium,golubov18}.

Under the assumption of uniform rotation about the maximum moment of inertia of an asteroid, the secular dynamics of the asteroid's rotation rate $\omega$ and obliquity $\varepsilon$ is described by \citep{rubincam00}:
\begin{align}
I_z \frac{\mathrm{d}\omega}{\mathrm{d}t}=T_z ,\label{domega-dt}\\
I_z \frac{\mathrm{d}\varepsilon}{\mathrm{d}t}=\frac{1}{\omega}T_\varepsilon.\label{deps-dt}
\end{align}
Here $I_z$ is the asteroid's moment of inertia, while $T_\omega$ and $T_\varepsilon$ are the axial and obliquity components of the mean YORP torque, acting on the asteroid, and $t$ is time.

The YORP torque creates a phase flow in the $\omega-\varepsilon$ plane, to which all asteroids are subject. Understanding the basic properties of this phase flow constitutes the most important problem of the theory of YORP. Without a good understanding of the topology of the asteroid's evolutionary trajectories, one cannot tackle such higher-level problems as distribution of asteroids over rotation rates and obliquities, understanding of evolutionary significance of tumbling, collisions, landslides, binary formation etc.

This paper is focused on the foundation of such a general description of the YORP evolution. We use both general theoretical considerations and numeric simulations of specific asteroids to single out the most generic types of evolutionary behavior, as well as to classify the possible stable equilibria where evolution can stall.

In Section \ref{sec:YORP without thermal inertia} we study the simplest model of the YORP torque, which neglects the thermal inertia and the tangential YORP. Under such assumptions, the YORP torques $T_\omega$ and $T_\varepsilon$ depend only on the obliquity $\varepsilon$, but not on the rotation rate $\omega$. We find that in most asteroids simple trigonometric functions of obliquity $\varepsilon$ can fit the simulated YORP very well. Moreover, YORP is described by one single parameter, and both axial and obliquity components of YORP for any obliquity can be expressed through this parameter. Observed correlations between YORP effects of different asteroids are seen to be very good, and a simple theoretic explanation exists for this correlation. 

The derived trigonometric fits to the YORP torque are used in Section \ref{sec:Dynamical evolution} to develop analytic expression for evolutionary tracks of asteroids subject to YORP. The resulting generic evolutionary tracks start as slow rotators, reach high rotation rates, and then either get disrupted by centrifugal forces, or return back to slow rotation.

In Section \ref{sec:equilibria} we add the tangential YORP as a new ingredient to our model. Then stable equilibria between the normal YORP and the tangential YORP become possible for several per cent of asteroids. These equilibria can work as sinks for asteroid rotation rates halting their overall spin evolution.

\section{YORP coefficients}
\label{sec:YORP without thermal inertia}

For the YORP torques, we choose the following approximations:
\begin{align}
T_z=\frac{\Phi R^3}{c}C_z (\cos{2\varepsilon}+\beta),\label{T_z_cos}\\
T_\varepsilon=\frac{\Phi R^3}{c}C_\varepsilon \sin{2\varepsilon}.\label{T_eps_sin}
\end{align}
Here $\Phi$ is the solar energy flux at the asteroid's orbit, $c$ is the speed of light, $R$ is the asteroid's mean radius, and their dimensional combination $\frac{\Phi R^3}{c}$ provides a scaling for the YORP-effect and roughly corresponds to an upper bound on YORP for an extremely asymmetric asteroid.
Next, $C_z$ and $C_\varepsilon$ are the dimensionless YORP coefficients, which are determined by the shape of the asteroid and are generally larger for more asymmetric asteroids, and $\beta$ is another fitting coefficient. Lastly, sine and cosine functions give an approximate dependence of $T_z$ and $T_\varepsilon$ on the obliquity $\varepsilon$. This particular obliquity dependence of YORP is chosen for its correct symmetry properties and for its general similarity to the obliquity dependence of YORP for many asteroids \citep{rubincam00,vokrouhlicky02}. Moreover, Eqn. (\ref{T_z_cos}) with $\beta=\frac{1}{3}$ follows from simplified theoretical models of YORP \citep{nesvorny07}, while Eqn. (\ref{T_eps_sin}) with $C_\varepsilon=\frac{2}{3}C_z$ represents the zero heat conductivity limit of the YORP theory by \cite{mysen08}. Note, that we assume no dependence of $T_z$ and $T_\varepsilon$ on the rotation rate. It holds for the axial component of the normal YORP $T_z$ in all cases \citep{golubov16nyorp}, while for the obliquity component $T_\varepsilon$ it is true only if the thermal inertia of the body can be neglected.

We compute the YORP torques for three different sets of asteroid shapes (photometric, radar and \textit{in situ}, see Appendix \ref{app:models}). For the computation we use the formalism of \cite{golubov16nyorp}, which was developed for convex shapes. Thus, for non-convex radar and \textit{in situ} models the computed YORP torques are approximate. The obtained torques as functions of obliquity are fitted by Eqs. (\ref{T_z_cos}) and (\ref{T_eps_sin}), using $C_z$, $C_\varepsilon$ and $\beta$ as free parameters. In most cases, the trigonometric fit indeed works well. Quantitatively, we characterize fineness of the fit by the parameter $\delta$, which is determined as the mean squared discrepancy between the computed YORP and the fit, normalized over the maximum of the computed YORP (see the Appendix \ref{app:models} for more detail). As an additional qualitative measure of the fit, we follow the classification by \cite{vokrouhlicky02}, who proposed to call asteroids type I and II, if the equation $T_\varepsilon(\varepsilon)=0$ has no roots at $0^\circ<\varepsilon<90^\circ$, and type III and IV, if the equation has such roots.\footnote{The distinction between classes I and II, as well as between classes III and IV, is unsubstantial. This distinction is determined via the sign of $T_\varepsilon$ at small positive $\varepsilon$. But interchange of the north and the south pole changes the signs of $T_\varepsilon$. Hereon, we will ignore this terminological distinction, and separate the asteroids into only two types, I/II and III/IV.} Naturally, type I/II has a better agreement with the trigonometric fit, than type III/IV. We find, that the trigonometric fit gives a good description for the obliquity dependence of YORP for the majority of asteroids, with about 70\% of asteroids simultaneously belonging to type I/II and having $\delta<0.4$. 

\begin{figure}
	\centering
	\includegraphics[width=0.48\textwidth]{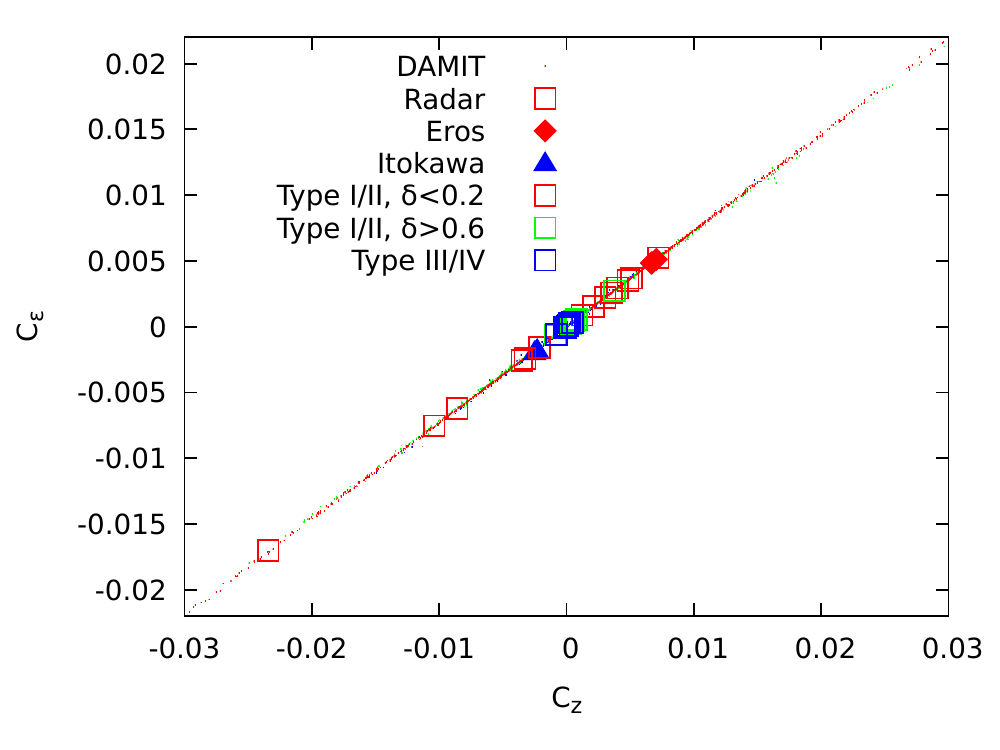}
	\caption{Correlation between the YORP coefficients $C_z$ and $C_\varepsilon$. Different points and symbols correspond to individual asteroids. Note, that each asteroid with \textit{in situ} shape determination is repeated four times and produces a flock of closely positioned points, as four models of different spatial resolution are treated separately. Fineness of trigonometric fit to YORP torques (Eqs. (\ref{T_z_cos}) and (\ref{T_eps_sin})) is color-coded. One sees a tight linear correlation between $C_z$ and $C_\varepsilon$. Another prominent trend is generally better fineness of fit for asteroids with greater YORP.}
	\label{fig:C_z-C_eps}
\end{figure}

\begin{figure}
	\centering
	\includegraphics[width=0.48\textwidth]{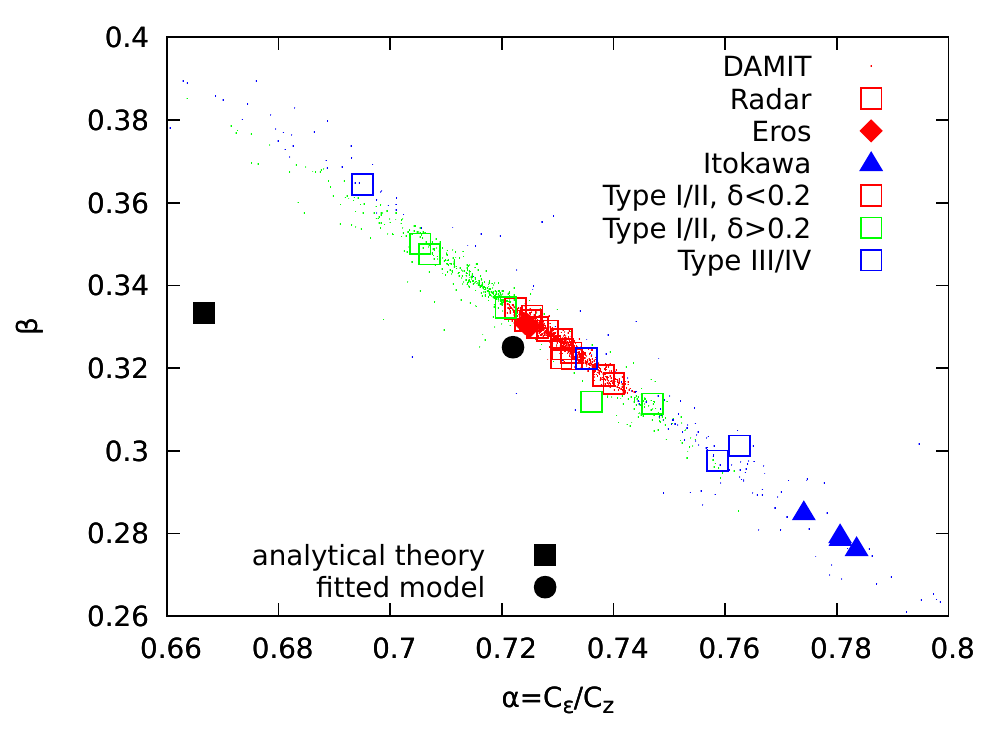}
	\caption{Correlation between the YORP coefficients $\alpha=C_\varepsilon/C_z$ and $\beta$. Markings are similar to Figure \ref{fig:C_z-C_eps}. Black square and circle mark theoretical predictions.}
	\label{fig:alpha-beta}
\end{figure}

When we plot the best-fit values for the axial YORP coefficient $C_z$ and the obliquity YORP coefficient $C_\varepsilon$ in Figure \ref{fig:C_z-C_eps}, we see a strikingly precise proportionality between them. This proportionality implies that $\alpha=C_\varepsilon/C_z$ should be constant to a good accuracy. In Figure \ref{fig:alpha-beta} we see, that both $\alpha$ and $\beta$ are constant to the accuracy of a few per cent.

The values of $\alpha$ and $\beta$ lie close to $\alpha=\frac{2}{3}$ and $\beta=\frac{1}{3}$, as expected from \cite{nesvorny07}, \cite{mysen08} and \cite{cicalo10}, although with some 10\% deviation in the case of $\alpha$ (black square in Figure \ref{fig:alpha-beta}).

The observed correlations can be even better explained in the framework of the analytic theory of YORP by \cite{golubov16nyorp}.
Performing the Taylor decomposition of analytic expressions for YORP (see Appendix \ref{app:taylor}), we arrive at Eqs. (\ref{T_z_cos}) and (\ref{T_eps_sin}), with $\beta=\frac{1}{3}$, and $C_z$ and $C_\varepsilon$ containing the exactly same integral over the asteroid surface, so that again $\alpha=C_\varepsilon/C_z=\frac{2}{3}$. The least-square-fit of sinusoidal laws to the theoretical expressions, which is more precise than the Taylor decomposition (see Appendix \ref{app:lsf}), gives a less straightforward but a more precise value $\alpha=0.722$, $\beta=0.325$, which agrees much better with the simulations (black circle in Figure \ref{fig:alpha-beta}).

Given the approximate invariance of $\alpha$ and $\beta$, the YORP effect of each asteroid is to a good accuracy characterized by a single coefficient $C_\varepsilon$, from which follow both components of YORP for any obliquity.

\section{Dynamical evolution of asteroids}
\label{sec:Dynamical evolution}

\begin{figure}
	\centering
	\begin{tikzpicture}[domain=0:8]
	\shade[shading=axis,bottom color=white,top color=blue!50,shading angle=180] (0,0.3) rectangle (8,0);
	\shade[shading=axis,bottom color=white,top color=blue!50,shading angle=0] (0,-0.3) rectangle (8,0);
	\shade[shading=axis,bottom color=white,top color=red!50,shading angle=0] (0,2.2) rectangle (8,2.7);
	\shade[shading=axis,bottom color=white,top color=red!50,shading angle=180] (0,-2.2) rectangle (8,-2.7);
	
	\draw[->] (0,0) -- (8.2,0);
	\draw[->] (0,-2.7) -- (0,2.7) node[left] {$\omega$};
	\draw[] (4,-2.7) -- (4,2.7);
	\draw[] (8,-2.7) -- (8,2.7);
	\draw[samples=50,domain=0:3.7]   plot (\x,{abs(sin(\x r * 3.1416/8))^0.92*abs(cos(\x r * 3.1416/8))^0.46});
	\draw[samples=51,domain=3.7:4.3]   plot (\x,{abs(sin(\x r * 3.1416/8))^0.92*abs(cos(\x r * 3.1416/8))^0.46});
	\draw[samples=50,domain=4.3:8]   plot (\x,{abs(sin(\x r * 3.1416/8))^0.92*abs(cos(\x r * 3.1416/8))^0.46});
	\draw[samples=50,domain=0:3.7]   plot (\x,{-abs(sin(\x r * 3.1416/8))^0.92*abs(cos(\x r * 3.1416/8))^0.46});
	\draw[samples=51,domain=3.7:4.3]   plot (\x,{-abs(sin(\x r * 3.1416/8))^0.92*abs(cos(\x r * 3.1416/8))^0.46});
	\draw[samples=50,domain=4.3:8]   plot (\x,{-abs(sin(\x r * 3.1416/8))^0.92*abs(cos(\x r * 3.1416/8))^0.46});
	\draw[samples=50,domain=0:3.7]   plot (\x,{2*abs(sin(\x r * 3.1416/8))^0.92*abs(cos(\x r * 3.1416/8))^0.46});
	\draw[samples=51,domain=3.7:4.3]   plot (\x,{2*abs(sin(\x r * 3.1416/8))^0.92*abs(cos(\x r * 3.1416/8))^0.46});
	\draw[samples=50,domain=4.3:8]   plot (\x,{2*abs(sin(\x r * 3.1416/8))^0.92*abs(cos(\x r * 3.1416/8))^0.46});
	\draw[samples=50,domain=0:3.7]   plot (\x,{-2*abs(sin(\x r * 3.1416/8))^0.92*abs(cos(\x r * 3.1416/8))^0.46});
	\draw[samples=51,domain=3.7:4.3]   plot (\x,{-2*abs(sin(\x r * 3.1416/8))^0.92*abs(cos(\x r * 3.1416/8))^0.46});
	\draw[samples=50,domain=4.3:8]   plot (\x,{-2*abs(sin(\x r * 3.1416/8))^0.92*abs(cos(\x r * 3.1416/8))^0.46});
	\draw[samples=50,domain=0:3.7]   plot (\x,{3*abs(sin(\x r * 3.1416/8))^0.92*abs(cos(\x r * 3.1416/8))^0.46});
	\draw[samples=51,domain=3.7:4.3]   plot (\x,{3*abs(sin(\x r * 3.1416/8))^0.92*abs(cos(\x r * 3.1416/8))^0.46});
	\draw[samples=50,domain=4.3:8]   plot (\x,{3*abs(sin(\x r * 3.1416/8))^0.92*abs(cos(\x r * 3.1416/8))^0.46});
	\draw[samples=50,domain=0:3.7]   plot (\x,{-3*abs(sin(\x r * 3.1416/8))^0.92*abs(cos(\x r * 3.1416/8))^0.46});
	\draw[samples=51,domain=3.7:4.3]   plot (\x,{-3*abs(sin(\x r * 3.1416/8))^0.92*abs(cos(\x r * 3.1416/8))^0.46});
	\draw[samples=50,domain=4.3:8]   plot (\x,{-3*abs(sin(\x r * 3.1416/8))^0.92*abs(cos(\x r * 3.1416/8))^0.46});
	\draw[samples=50,domain=0:3.7]   plot (\x,{4*abs(sin(\x r * 3.1416/8))^0.92*abs(cos(\x r * 3.1416/8))^0.46});
	\draw[samples=51,domain=3.7:4.3]   plot (\x,{4*abs(sin(\x r * 3.1416/8))^0.92*abs(cos(\x r * 3.1416/8))^0.46});
	\draw[samples=50,domain=4.3:8]   plot (\x,{4*abs(sin(\x r * 3.1416/8))^0.92*abs(cos(\x r * 3.1416/8))^0.46});
	\draw[samples=50,domain=0:3.7]   plot (\x,{-4*abs(sin(\x r * 3.1416/8))^0.92*abs(cos(\x r * 3.1416/8))^0.46});
	\draw[samples=51,domain=3.7:4.3]   plot (\x,{-4*abs(sin(\x r * 3.1416/8))^0.92*abs(cos(\x r * 3.1416/8))^0.46});
	\draw[samples=50,domain=4.3:8]   plot (\x,{-4*abs(sin(\x r * 3.1416/8))^0.92*abs(cos(\x r * 3.1416/8))^0.46});

	\draw[->]  (2.43,2.58)--(2.44,2.58);
	\draw[->]  (2.43,1.935)--(2.44,1.935);
	\draw[->]  (2.43,1.29)--(2.44,1.29);
	\draw[->]  (2.43,0.645)--(2.44,0.645);
	\draw[->]  (2.44,-2.58)--(2.43,-2.58);
	\draw[->]  (2.44,-1.935)--(2.43,-1.935);
	\draw[->]  (2.44,-1.29)--(2.43,-1.29);
	\draw[->]  (2.44,-0.645)--(2.43,-0.645);
	\draw[->]  (5.57,2.58)--(5.56,2.58);
	\draw[->]  (5.57,1.935)--(5.56,1.935);
	\draw[->]  (5.57,1.29)--(5.56,1.29);
	\draw[->]  (5.57,0.645)--(5.56,0.645);
	\draw[->]  (5.56,-2.58)--(5.57,-2.58);
	\draw[->]  (5.56,-1.935)--(5.57,-1.935);
	\draw[->]  (5.56,-1.29)--(5.57,-1.29);
	\draw[->]  (5.56,-0.645)--(5.57,-0.645);
	
	\fill[gray!50] (0,3.5) ellipse (0.3 and 0.3);
	\fill[white] (0,3.5) ellipse (0.3 and 0.12);
	\draw[] (0,3.5) ellipse (0.3 and 0.12);
	\fill[white] (-0.3,3.5) arc(180:0:0.3) -- cycle;
	\draw[] (0,3.5) ellipse (0.3 and 0.3);
	\begin{scope}
	\path[clip] (0.07,3.95)--(0.07,4.05)--(0.25,4.05)--(0.25,3.75)--(-0.25,3.75)--(-0.25,4.05)--(-0.07,4.05)--(-0.07,3.95)--cycle;
	\draw[] (0,3.95) ellipse (0.2 and 0.08);
	\end{scope}
	\draw[->]  (0.073,4.03)--(0.07,4.031);
	\draw[] (0,3.75) -- (0,3.84);
	\draw[] (0,3.9) -- (0,4.1);
	\begin{scope}
	\path[clip] (0.07,3.05)--(0.07,3.15)--(0.25,3.15)--(0.25,2.95)--(-0.25,2.95)--(-0.25,3.15)--(-0.07,3.15)--(-0.07,3.05)--cycle;
	\draw[] (0,3.05) ellipse (0.2 and 0.08);
	\end{scope}
	\draw[->]  (0.073,3.13)--(0.07,3.131);
	\draw[] (0,3.2) -- (0,3.0);
	\draw[] (0,2.94) -- (0,2.9);
	
	\fill[white] (8,3.5) ellipse (0.3 and 0.3);
	\fill[gray!50] (8,3.5) ellipse (0.3 and 0.12);
	\draw[] (8,3.5) ellipse (0.3 and 0.12);
	\fill[gray!50] (7.7,3.5) arc(180:0:0.3) -- cycle;
	\draw[] (8,3.5) ellipse (0.3 and 0.3);
	\begin{scope}
	\path[clip] (8.07,3.95)--(8.07,4.05)--(8.25,4.05)--(8.25,3.75)--(7.75,3.75)--(7.75,4.05)--(7.93,4.05)--(7.93,3.95)--cycle;
	\draw[] (8,3.95) ellipse (0.2 and 0.08);
	\end{scope}
	\draw[->]  (7.927,4.03)--(7.93,4.031);
	\draw[] (8,3.75) -- (8,3.84);
	\draw[] (8,3.9) -- (8,4.1);
	\begin{scope}
	\path[clip] (8.07,3.05)--(8.07,3.15)--(8.25,3.15)--(8.25,2.95)--(7.75,2.95)--(7.75,3.15)--(7.93,3.15)--(7.93,3.05)--cycle;
	\draw[] (8,3.05) ellipse (0.2 and 0.08);
	\end{scope}
	\draw[->]  (7.927,3.13)--(7.93,3.131);
	\draw[] (8,3.2) -- (8,3.0);
	\draw[] (8,2.94) -- (8,2.9);
	
	\fill[gray!50] (4,3.5) ellipse (0.3 and 0.3);
	\fill[white] (4,3.5) ellipse (0.12 and 0.3);
	\draw[] (4,3.5) ellipse (0.12 and 0.3);
	\fill[white] (4,3.8) arc(90:-90:0.3) -- cycle;
	\draw[] (4,3.5) ellipse (0.3 and 0.3);
	\begin{scope}
	\path[clip] (4.45,3.57)--(4.55,3.57)--(4.55,3.75)--(4.25,3.75)--(4.25,3.25)--(4.55,3.25)--(4.55,3.43)--(4.45,3.43)--cycle;
	\draw[] (4.45,3.5) ellipse (0.08 and 0.2);
	\end{scope}
	\draw[->]  (4.53,3.427)--(4.531,3.43);
	\draw[] (4.25,3.5) -- (4.35,3.5);
	\draw[] (4.39,3.5) -- (4.6,3.5);
	\begin{scope}
	\path[clip] (3.55,3.57)--(3.75,3.57)--(3.75,3.75)--(3.45,3.75)--(3.45,3.25)--(3.75,3.25)--(3.75,3.43)--(3.55,3.43)--cycle;
	\draw[] (3.55,3.5) ellipse (0.08 and 0.2);
	\end{scope}
	\draw[->]  (3.629,3.427)--(3.63,3.43);
	\draw[] (3.7,3.5) -- (3.49,3.5);
	\draw[] (3.45,3.5) -- (3.4,3.5);
	
	\fill[gray!50] (0,-3.5) ellipse (0.3 and 0.3);
	\fill[white] (0,-3.5) ellipse (0.3 and 0.12);
	\draw[] (0,-3.5) ellipse (0.3 and 0.12);
	\fill[white] (-0.3,-3.5) arc(180:0:0.3) -- cycle;
	\draw[] (0,-3.5) ellipse (0.3 and 0.3);
	\begin{scope}
	\path[clip] (0.07,-3.95)--(0.07,-3.85)--(0.25,-3.85)--(0.25,-4.05)--(-0.25,-4.05)--(-0.25,-3.85)--(-0.07,-3.85)--(-0.07,-3.95)--cycle;
	\draw[] (0,-3.95) ellipse (0.2 and 0.08);
	\end{scope}
	\draw[->]  (-0.073,-3.871)--(-0.07,-3.87);
	\draw[] (0,-3.8) -- (0,-4.0);
	\draw[] (0,-4.06) -- (0,-4.1);
	\begin{scope}
	\path[clip] (0.07,-3.05)--(0.07,-2.95)--(0.25,-2.95)--(0.25,-3.15)--(-0.25,-3.15)--(-0.25,-2.95)--(-0.07,-2.95)--(-0.07,-3.05)--cycle;
	\draw[] (0,-3.05) ellipse (0.2 and 0.08);
	\end{scope}
	\draw[->]  (-0.073,-2.971)--(-0.07,-2.97);
	\draw[] (0,-3.25) -- (0,-3.16);
	\draw[] (0,-3.1) -- (0,-2.9);
	
	\fill[white] (8,-3.5) ellipse (0.3 and 0.3);
	\fill[gray!50] (8,-3.5) ellipse (0.3 and 0.12);
	\draw[] (8,-3.5) ellipse (0.3 and 0.12);
	\fill[gray!50] (7.7,-3.5) arc(180:0:0.3) -- cycle;
	\draw[] (8,-3.5) ellipse (0.3 and 0.3);
	\begin{scope}
	\path[clip] (8.07,-3.95)--(8.07,-3.85)--(8.25,-3.85)--(8.25,-4.05)--(7.75,-4.05)--(7.75,-3.85)--(7.93,-3.85)--(7.93,-3.95)--cycle;
	\draw[] (8,-3.95) ellipse (0.2 and 0.08);
	\end{scope}
	\draw[->]  (8.077,-3.871)--(8.07,-3.87);
	\draw[] (8,-3.8) -- (8,-4.0);
	\draw[] (8,-4.06) -- (8,-4.1);
	\begin{scope}
	\path[clip] (8.07,-3.05)--(8.07,-2.95)--(8.25,-2.95)--(8.25,-3.15)--(7.75,-3.15)--(7.75,-2.95)--(7.93,-2.95)--(7.93,-3.05)--cycle;
	\draw[] (8,-3.05) ellipse (0.2 and 0.08);
	\end{scope}
	\draw[->]  (8.073,-2.971)--(8.07,-2.97);
	\draw[] (8,-3.25) -- (8,-3.16);
	\draw[] (8,-3.1) -- (8,-2.9);
	
	\fill[gray!50] (4,-3.5) ellipse (0.3 and 0.3);
	\fill[white] (4,-3.5) ellipse (0.12 and 0.3);
	\draw[] (4,-3.5) ellipse (0.12 and 0.3);
	\fill[white] (4,-3.8) arc(-90:90:0.3) -- cycle;
	\draw[] (4,-3.5) ellipse (0.3 and 0.3);
	\begin{scope}
	\path[clip] (4.45,-3.57)--(4.55,-3.57)--(4.55,-3.75)--(4.25,-3.75)--(4.25,-3.25)--(4.55,-3.25)--(4.55,-3.43)--(4.45,-3.43)--cycle;
	\draw[] (4.45,-3.5) ellipse (0.08 and 0.2);
	\end{scope}
	\draw[->]  (4.53,-3.427)--(4.531,-3.43);
	\draw[] (4.25,-3.5) -- (4.35,-3.5);
	\draw[] (4.39,-3.5) -- (4.6,-3.5);
	\begin{scope}
	\path[clip] (3.55,-3.57)--(3.75,-3.57)--(3.75,-3.75)--(3.45,-3.75)--(3.45,-3.25)--(3.75,-3.25)--(3.75,-3.43)--(3.55,-3.43)--cycle;
	\draw[] (3.55,-3.5) ellipse (0.08 and 0.2);
	\end{scope}
	\draw[->]  (3.629,-3.427)--(3.63,-3.43);
	\draw[] (3.7,-3.5) -- (3.49,-3.5);
	\draw[] (3.45,-3.5) -- (3.4,-3.5);
	
	\node[] at (8.2,0.2) {$\varepsilon$};
	\node[blue] at (6,-0.2) {tumbling};
	\node[red] at (7.1,2.5) {disruption};
	\node[] at (-0.1,0) {0};
	\node[] at (4.3,0.2) {$90^\circ$};
	\node[] at (8.2,-0.2) {$180^\circ$};
	
	\end{tikzpicture}
	\caption{Evolution diagram for a simple evolutionary model. Rotation rate $\omega$ and obliquity $\varepsilon$ are plotted along the coordinate axes. Small icons illustrate the orientation and the direction of rotation of the asteroid. Black lines with arrows show evolutionary tracks of asteroids with different initial conditions. The disruption limit and the tumbling region are shown in red and blue respectively.
	}
	\label{model-simplest}
\end{figure}
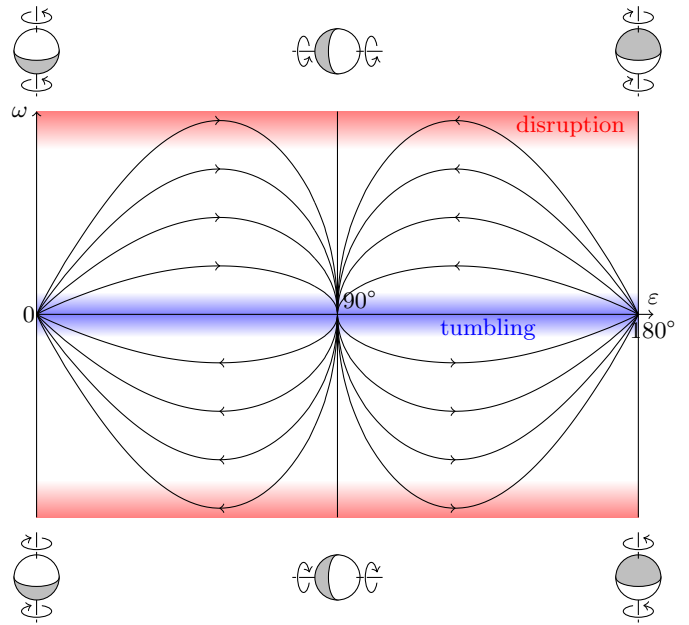

Substituting approximate analytic expressions for YORP from Eqs. (\ref{T_z_cos}) and (\ref{T_eps_sin}) into general evolutionary equations Eqs. (\ref{domega-dt}) and (\ref{deps-dt}), we get a closed set of equations that specifies evolution of the asteroid spin state. 

Eliminating time from this set of equations, we get a separable equation for $\omega$ and $\varepsilon$. Its solution describes integral curves of the phase flow,
\begin{equation}
\omega=\omega_{0}\left(\sin{\varepsilon}\right)^{\frac{1+\beta}{2\alpha}}\left(\cos{\varepsilon}\right)^{\frac{1-\beta}{2\alpha}}.
\label{omega_eps}\\
\end{equation}
Here $\omega_{0}$ is the integration constant. This equation generalizes the first integral reported by \cite{cicalo10}. The maximal rotation rate in Eqn. (\ref{omega_eps}) is reached at $\varepsilon=\frac{1}{2}\arccos(-\beta)=54.5^\circ$ and equals $\omega_\mathrm{max}=0.64\omega_0$. 

The phase curves are illustrated in Figure \ref{model-simplest}. Phase trajectories given by Eqn. (\ref{omega_eps}) are plotted with black lines, with the arrows marking the direction of evolution. 
The plot shows the area $0^\circ<\varepsilon<180^\circ$, $-\infty<\omega<\infty$. The plot is symmetric with respect to the transformation $\varepsilon\rightarrow 180^\circ-\varepsilon$. Small icons show the orientation and the rotation of the asteroid at six different positions in the plot, with the northern and the southern hemispheres of the asteroid shown in white and grey respectively. Bear in mind, that we determine the north pole on the asteroid by requiring $C_\varepsilon>0$. 

The red area in the plot marks the disruption limit of the asteroids. The blue area marks the region of slow rotators subject to tumbling. (The relative size of the tumbling region is exaggerated.) The theory presented in this paper is applicable only in the region between the blue and red areas.

All phase curves emerge from the tumbling regime and either lead the asteroid to disruption, or return back to tumbling. 
For $\omega>0$ the motion along these phase curves starts at $\varepsilon=0^\circ$ or $\varepsilon=180^\circ$, and goes to $\varepsilon=90^\circ$. For $\omega<0$ the motion starts at $\varepsilon=90^\circ$ and goes to $\varepsilon=0^\circ$ or $\varepsilon=180^\circ$.

Substituting the equation of the phase trajectory Eqn. (\ref{omega_eps}) back into Eqs. (\ref{domega-dt}) and (\ref{deps-dt}), one can derive temporal dependencies $\omega(t)$ and $\varepsilon(t)$, although not in elementary functions. Let us only mention that the whole cycle from tumbling via high rotation rates and back to tumbling takes a time
\begin{equation}
t_0=1.43\frac{cI_z\omega_0}{\Phi R^3C_\varepsilon}.
\end{equation}
As $I_z\propto R^5$, we arrive at the usual scaling law for the YORP evolution timescale $t_0\propto R^2$.

\section{YORP equilibria}
\label{sec:equilibria}
The phase flow discussed above has simple self-similar trajectories and a trivial topology. Still, more complex evolution of asteroid spins will occur, if one takes into consideration the tangential YORP (or TYORP).

TYORP is caused by asymmetric light emission by boulders or other structures on the surface of the asteroid \cite{golubov12,golubov17}. It tends to accelerate the asteroid spin rate rather than to decelerate it in all known cases, and to be maximal at some particular rotation rate, while decreasing for greater and smaller rotation rates.

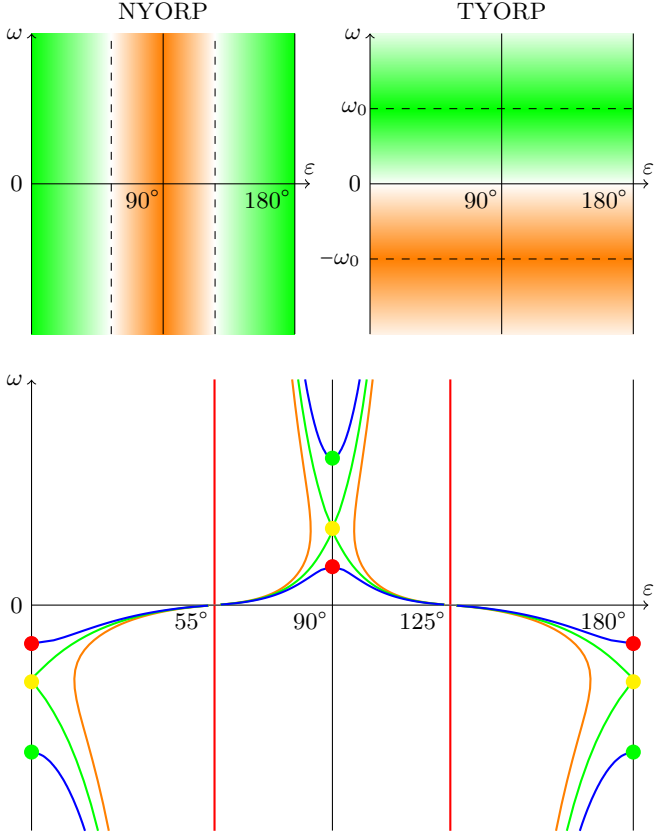
\begin{figure}
	\centering
	\begin{tikzpicture}[domain=0:8]
	\shade[shading=axis,top color=green,bottom color=white,shading angle=90] (0,-2) rectangle (1.06,2);
	\shade[shading=axis,top color=white,bottom color=orange,shading angle=90] (1.06,-2) rectangle (1.75,2);
	\shade[shading=axis,top color=orange,bottom color=white,shading angle=90] (1.75,-2) rectangle (2.44,2);
	\shade[shading=axis,top color=white,bottom color=green,shading angle=90] (2.44,-2) rectangle (3.5,2);
	
	\draw[->] (0,0) -- (3.7,0);
	\draw[->] (0,-2) -- (0,2) node[left] {$\omega$};
	\draw[] (1.75,-2) -- (1.75,2);
	\draw[] (3.5,-2) -- (3.5,2);
	\draw[dashed] (1.06,-2) -- (1.06,2);
	\draw[dashed] (2.44,-2) -- (2.44,2);
	
	\node[] at (3.7,0.2) {$\varepsilon$};
	\node[] at (-0.2,0) {0};
	\node[] at (1.48,-0.2) {$90^\circ$};
	\node[] at (3.14,-0.2) {$180^\circ$};
	
	\shade[shading=axis,top color=white,bottom color=green,shading angle=180] (4.5,0) rectangle (8,1);
	\shade[shading=axis,top color=green,bottom color=green!10,shading angle=180] (4.5,1) rectangle (8,2);
	\shade[shading=axis,top color=white,bottom color=orange,shading angle=0] (4.5,0) rectangle (8,-1);
	\shade[shading=axis,top color=orange,bottom color=orange!10,shading angle=0] (4.5,-1) rectangle (8,-2);
		
	\draw[->] (4.5,0) -- (8.2,0);
	\draw[->] (4.5,-2) -- (4.5,2) node[left] {$\omega$};
	\draw[] (6.25,-2) -- (6.25,2);
	\draw[] (8,-2) -- (8,2);
	\draw[dashed] (4.5,1) -- (8,1);
	\draw[dashed] (4.5,-1) -- (8,-1);
	
	\node[] at (8.2,0.2) {$\varepsilon$};
	\node[] at (4.3,1) {$\omega_0$};
	\node[] at (4.1,-1) {$-\omega_0$};
	\node[] at (4.3,0) {0};
	\node[] at (5.98,-0.2) {$90^\circ$};
	\node[] at (7.62,-0.2) {$180^\circ$};
	
	\node[] at (1.75,2.3) {NYORP};
	\node[] at (6.25,2.3) {TYORP};
	
	\end{tikzpicture}
	
	\begin{tikzpicture}[domain=0:8]
	\draw[->] (0,0) -- (8.2,0);
	\draw[->] (0,-3) -- (0,3) node[left] {$\omega$};
	\draw[] (4,-3) -- (4,3);
	\draw[] (8,-3) -- (8,3);
	
	\node[] at (8.2,0.2) {$\varepsilon$};
	\node[] at (-0.2,0) {0};
	\node[] at (3.71,-0.2) {$90^\circ$};
	\node[] at (7.61,-0.2) {$180^\circ$};
	\node[] at (2.13,-0.2) {$55^\circ$};
	\node[] at (5.2,-0.2) {$125^\circ$};
	
	\draw[thick,red] plot [domain=-5:1.1,samples=50] ({8/3.1416*0.5*rad(acos(-(0*1.5*2.71828^(-\x*\x/4/1.518/1.518)+1./3.)/(1+0*0.5*2.71828^(-\x*\x/4/1.518/1.518))))},{2.71828^(\x)});
	\draw[thick,orange] plot [domain=-5:1.1,samples=50] ({8/3.1416*0.5*rad(acos(-(0.633*1.5*2.71828^(-\x*\x/4/1.518/1.518)+1./3.)/(1+0.633*0.5*2.71828^(-\x*\x/4/1.518/1.518))))},{2.71828^(\x)});
	\draw[thick,green] plot [domain=-5:1.1,samples=50] ({8/3.1416*0.5*rad(acos(-(0.6666*1.5*2.71828^(-\x*\x/4/1.518/1.518)+1./3.)/(1+0.6666*0.5*2.71828^(-\x*\x/4/1.518/1.518))))},{2.71828^(\x)});
	\draw[thick,blue] plot [domain=0.67061:1.1,samples=50] ({8/3.1416*0.5*rad(acos(-(0.7*1.5*2.71828^(-\x*\x/4/1.518/1.518)+1./3.)/(1+0.7*0.5*2.71828^(-\x*\x/4/1.518/1.518))))},{2.71828^(\x)});
	\draw[thick,blue] plot [domain=-5:-0.67061,samples=50] ({8/3.1416*0.5*rad(acos(-(0.7*1.5*2.71828^(-\x*\x/4/1.518/1.518)+1./3.)/(1+0.7*0.5*2.71828^(-\x*\x/4/1.518/1.518))))},{2.71828^(\x)});
	
	\draw[thick,red] plot [domain=-5:1.1,samples=50] ({8-8/3.1416*0.5*rad(acos(-(0*1.5*2.71828^(-\x*\x/4/1.518/1.518)+1./3.)/(1+0*0.5*2.71828^(-\x*\x/4/1.518/1.518))))},{2.71828^(\x)});
	\draw[thick,orange] plot [domain=-5:1.1,samples=50] ({8-8/3.1416*0.5*rad(acos(-(0.633*1.5*2.71828^(-\x*\x/4/1.518/1.518)+1./3.)/(1+0.633*0.5*2.71828^(-\x*\x/4/1.518/1.518))))},{2.71828^(\x)});
	\draw[thick,green] plot [domain=-5:1.1,samples=50] ({8-8/3.1416*0.5*rad(acos(-(0.6666*1.5*2.71828^(-\x*\x/4/1.518/1.518)+1./3.)/(1+0.6666*0.5*2.71828^(-\x*\x/4/1.518/1.518))))},{2.71828^(\x)});
	\draw[thick,blue] plot [domain=0.67061:1.1,samples=50] ({8-8/3.1416*0.5*rad(acos(-(0.7*1.5*2.71828^(-\x*\x/4/1.518/1.518)+1./3.)/(1+0.7*0.5*2.71828^(-\x*\x/4/1.518/1.518))))},{2.71828^(\x)});
	\draw[thick,blue] plot [domain=-5:-0.67061,samples=50] ({8-8/3.1416*0.5*rad(acos(-(0.7*1.5*2.71828^(-\x*\x/4/1.518/1.518)+1./3.)/(1+0.7*0.5*2.71828^(-\x*\x/4/1.518/1.518))))},{2.71828^(\x)});
	
	\draw[thick,red] plot [domain=-5:1.1,samples=50] ({8/3.1416*0.5*rad(acos((0*1.5*2.71828^(-\x*\x/4/1.518/1.518)-1./3.)/(1-0*0.5*2.71828^(-\x*\x/4/1.518/1.518))))},{-2.71828^(\x)});
	\draw[thick,orange] plot [domain=-5:1.1,samples=50] ({8/3.1416*0.5*rad(acos((0.633*1.5*2.71828^(-\x*\x/4/1.518/1.518)-1./3.)/(1-0.633*0.5*2.71828^(-\x*\x/4/1.518/1.518))))},{-2.71828^(\x)});
	\draw[thick,green] plot [domain=-5:1.1,samples=50] ({8/3.1416*0.5*rad(acos((0.6666*1.5*2.71828^(-\x*\x/4/1.518/1.518)-1./3.)/(1-0.6666*0.5*2.71828^(-\x*\x/4/1.518/1.518))))},{-2.71828^(\x)});
	\draw[thick,blue] plot [domain=0.67061:1.1,samples=50] ({8/3.1416*0.5*rad(acos((0.7*1.5*2.71828^(-\x*\x/4/1.518/1.518)-1./3.)/(1-0.7*0.5*2.71828^(-\x*\x/4/1.518/1.518))))},{-2.71828^(\x)});
	\draw[thick,blue] plot [domain=-5:-0.67061,samples=50] ({8/3.1416*0.5*rad(acos((0.7*1.5*2.71828^(-\x*\x/4/1.518/1.518)-1./3.)/(1-0.7*0.5*2.71828^(-\x*\x/4/1.518/1.518))))},{-2.71828^(\x)});
	
	\draw[thick,red] plot [domain=-5:1.1,samples=50] ({8-8/3.1416*0.5*rad(acos((0*1.5*2.71828^(-\x*\x/4/1.518/1.518)-1./3.)/(1-0*0.5*2.71828^(-\x*\x/4/1.518/1.518))))},{-2.71828^(\x)});
	\draw[thick,orange] plot [domain=-5:1.1,samples=50] ({8-8/3.1416*0.5*rad(acos((0.633*1.5*2.71828^(-\x*\x/4/1.518/1.518)-1./3.)/(1-0.633*0.5*2.71828^(-\x*\x/4/1.518/1.518))))},{-2.71828^(\x)});
	\draw[thick,green] plot [domain=-5:1.1,samples=50] ({8-8/3.1416*0.5*rad(acos((0.6666*1.5*2.71828^(-\x*\x/4/1.518/1.518)-1./3.)/(1-0.6666*0.5*2.71828^(-\x*\x/4/1.518/1.518))))},{-2.71828^(\x)});
	\draw[thick,blue] plot [domain=0.67061:1.1,samples=50] ({8-8/3.1416*0.5*rad(acos((0.7*1.5*2.71828^(-\x*\x/4/1.518/1.518)-1./3.)/(1-0.7*0.5*2.71828^(-\x*\x/4/1.518/1.518))))},{-2.71828^(\x)});
	\draw[thick,blue] plot [domain=-5:-0.67061,samples=50] ({8-8/3.1416*0.5*rad(acos((0.7*1.5*2.71828^(-\x*\x/4/1.518/1.518)-1./3.)/(1-0.7*0.5*2.71828^(-\x*\x/4/1.518/1.518))))},{-2.71828^(\x)});\textsc{}
	
	\fill[green] (4,1.9554) ellipse (0.1 and 0.1);
	\fill[yellow] (4,1.02) ellipse (0.1 and 0.1);
	\fill[red] (4,0.5114) ellipse (0.1 and 0.1);
	\fill[green] (0,-1.9554) ellipse (0.1 and 0.1);
	\fill[yellow] (0,-1.02) ellipse (0.1 and 0.1);
	\fill[red] (0,-0.5114) ellipse (0.1 and 0.1);
	\fill[green] (8,-1.9554) ellipse (0.1 and 0.1);
	\fill[yellow] (8,-1.02) ellipse (0.1 and 0.1);
	\fill[red] (8,-0.5114) ellipse (0.1 and 0.1);
	
	\end{tikzpicture}
	\caption{Evolution diagrams including the tangential YORP.
		\textit{Top:} The areas of increase and decrease of the rotation rate $\omega$ due to the normal YORP and the tangential YORP effect are marked with green and orange colors correspondingly, the denser colors for the bigger absolute value. \textit{Bottom:} Lines of zero angular acceleration $\dot{\omega}=0$ for cases of different relative strength of NYORP and TYORP. For no TYORP $\dot{\omega}=0$ is attained on a vertical straight line (red line). If a small amount of the tangential YORP is added to the normal YORP, the line bends (orange line). Then a bifurcation occurs (green line, yellow dot). At even stronger TYORP, the line of zero angular acceleration intersects the lines of zero obliquity acceleration ($\varepsilon=0,\,90^\circ,\,180^\circ$, and six equilibrium points appear, for which both angular and axial accelerations are 0 (blue line). The stable and unstable equilibria are marked with green and red dots respectively.}
	\label{fig:tyorp}
\end{figure}

The impact of TYORP on the dynamics of an asteroid is illustrated in Figure \ref{fig:tyorp}. If TYORP is large enough, six equilibrium points appear, two at each of the lines $\varepsilon=0^\circ$, $\varepsilon=90^\circ$ and $\varepsilon=180^\circ$. On each line one point is stable and the other is unstable. Interestingly, in the simple model all the three bifurcations creating the six equilibrium points occur simultaneously. It happens because in the simple model both NYORP and TYORP at $\varepsilon=0^\circ$ are exactly 2 times larger than at $90^\circ$.

\begin{figure}
	\centering
	\includegraphics[width=0.48\textwidth]{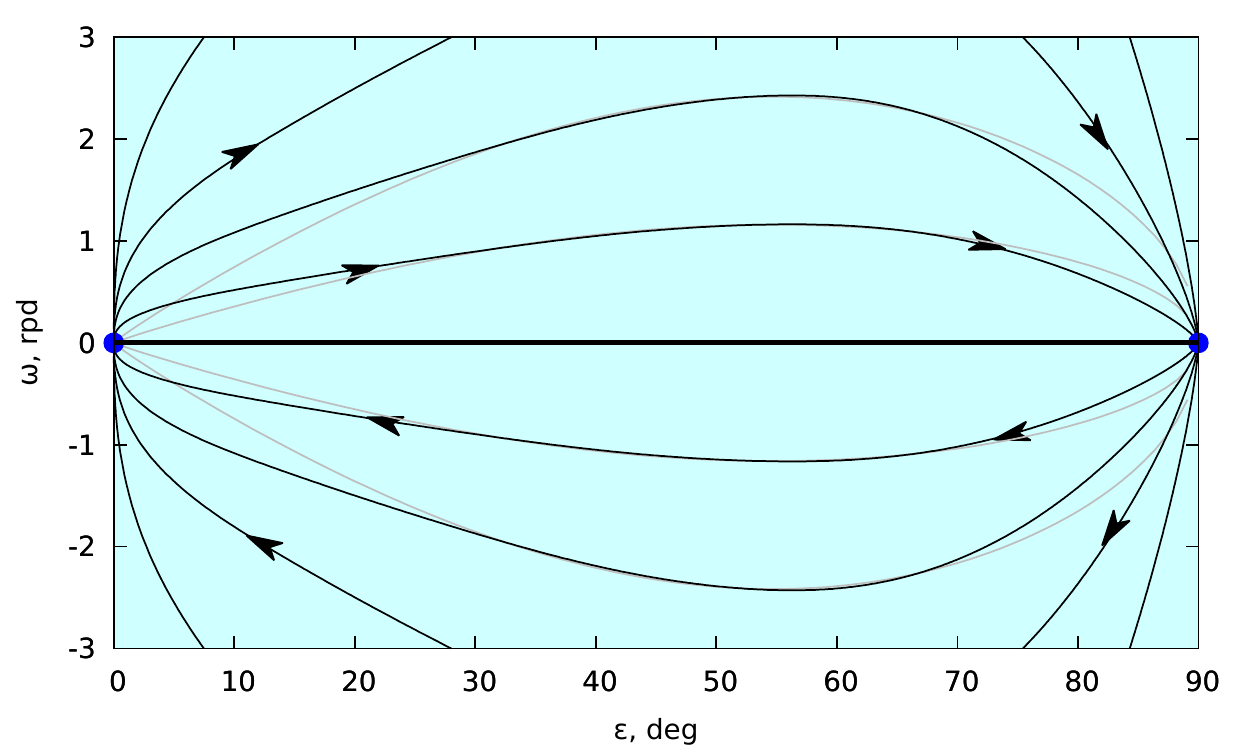}\\
	\includegraphics[width=0.48\textwidth]{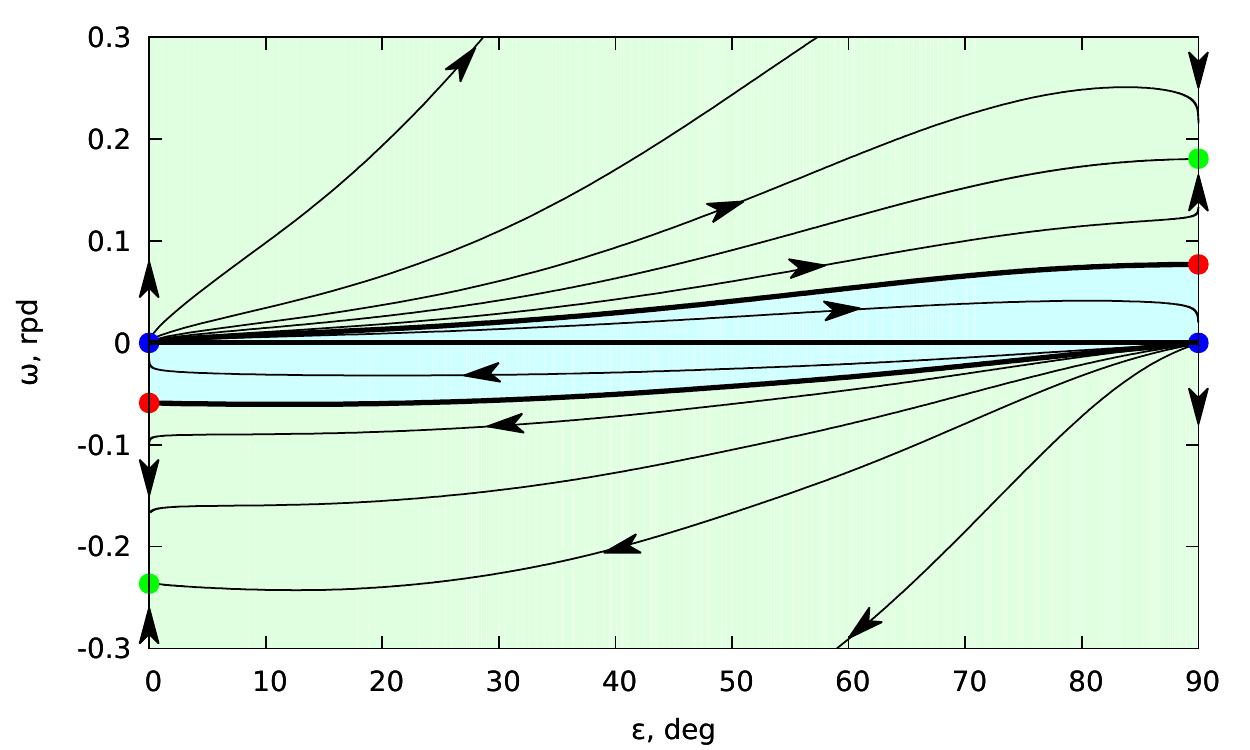}
	\caption{Evolution diagrams for asteroid 4660 Nereus. \textit{Top:} Zero TYORP is assumed, i.e. $n_0=0$. Evolutionary trend given by analytic expression Eqn. (\ref{omega_eps}) is overplotted with grey lines. \textit{Bottom:} TYORP corresponding to $n_0=0.09$. Stable equilibria are marked with green circles, unstable equilibria - with red dots, attractors at slow rotation - with blue dots. Basins of different attractors are shown in the colors of the attractors and separated by bold black lines.}
	\label{fig:tyorp-diagrams-nereus}
\end{figure}

Simulations of the evolution of an asteroid is illustrated in Figure \ref{fig:tyorp-diagrams-nereus}. Asteroid 4660 Nereus is taken as a typical example. It has $\delta=0.21$, thus the agreement between its trajectories for zero TYORP with Eqn. (\ref{omega_eps}) is moderate. If we add TYORP, we observe stable and unstable equilibria (compare Figure \ref{fig:tyorp} and the bottom panel of Figure \ref{fig:tyorp-diagrams-nereus}). 

The presence of stable equilibria can drastically change the spin dynamics of asteroids. They are no longer destined to head to either disruption or tumbling, thus repeating the YORP cycles until they ultimately decay, but can instead be attracted to stable equilibria, which serve as sinks and eliminate the asteroids from undergoing YORP cycles.

In Appendix \ref{app:equilibria}, we discuss conditions for existence of such equilibria in more detail, and find that they are very probable. The probability for a stable equilibrium to exist depends on the number of boulders on the surface, their thermal parameter, details of the asteroid shape models etc., and in most cases that we computed exceeds 10\%. 

The probability for an asteroid to be locked in such an equilibrium depends on the initial conditions, but if the asteroid in the course of its evolution changes its shape due to centrifugal forces (or a collisional event), then it has a substantial probability of finding itself within the attractor of a stable equilibrium. Therefore, each asteroid has a high chance of being attracted to such an equilibrium after a few YORP cycles.

Equilibria at $\varepsilon=0^\circ\, / \,180^\circ $ and $\varepsilon=90^\circ$ are approximately equally probable. The distribution over rotation rates is also biased towards smaller rotation rates, although at the slowest rotation rates (less than one revolution per several days for cracked stone and hence low thermal inertia) no equilibria are possible.

\section{Discussion}
\label{sec:Discussion}

We find that the YORP effect of most asteroids is well fitted by trigonometric functions of obliquity (Eqs. (\ref{T_z_cos}) and (\ref{T_eps_sin})). The study of the known asteroid shape models reveals a fundamental correlation, that we see exists across all asteroid shapes (Figure \ref{fig:C_z-C_eps}). This correlation follows from the Taylor decomposition of the YORP expression used in our computations. The observed correlations could be worsened by such effects disregarded by this expression, as shadowing or self-illumination on a concave asteroid surface or possible differences in scattering laws on different parts of the asteroid body. Note, that even simulating non-convex asteroid shapes, we did it with equations that were derived for convex bodies and demonstrated to work satisfactory for moderately non-convex shapes \citep{golubov16nyorp}. The minority of the asteroids, for which the proposed trigonometric fit does not work, remained beyond the scope of our study (type III/IV). The effect of thermal lag between the absorbed and the emitted heat on the obliquity evolution was also disregarded. All these issues present important directions for future work. 

The first-order trigonometric fit allows us to construct a family of evolutionary trajectories of an asteroid (Figure \ref{model-simplest}). This family of trajectories is universal within the simple model, although the shape and even the topology of the trajectories can be altered in more inclusive physical models, e. g. by including thermal lag or TYORP, or considering type III/IV asteroids. More universal is the $\varepsilon$-$\omega$ diagram itself, which presents the main battleground between different torques acting upon the asteroid. 

From the top and the bottom the diagram is limited by the disruption limit: if $|\omega|$ becomes too big, the shape and the dynamics of the asteroid can get altered by landslides, mass shedding, formation of a satellite and the subsequent gravitational interaction with the satellite. The landslides alone alter the YORP experienced by the asteroid, and this changes the geometry of the possible evolution curves for this asteroid \citep{harris09,statler09}.

Positive and negative $\omega$ in the diagram are separated by the region of tumbling. In this region, the two-dimensional system characterized by $\varepsilon$ and $\omega$, acquires a third dimension. A tumbler can be described by its energy $E$, angular momentum $L$ and the obliquity of the angular momentum with respect to the orbital plane $\varepsilon$. At large $L$, the body relaxes to the principal-axis rotation, then $L$ and $E$ get connected, $\omega$ stops continuously oscillating and can be expressed through either $L$ or $E$, and we return back to the two-dimensional description of the system via $\varepsilon$ and $\omega$. The third dimension, suppressed in the major part of the diagram, can play an important role in the tumbling region by resetting slow rotators to another part of the diagram.

To understand the evolution of asteroids, one must study the phase flow of asteroids in an ensemble of $\varepsilon$-$\omega$ diagrams corresponding to different shapes and other properties of the asteroid. The theory of YORP determines the geometry of this phase flow. Tumbling and disruption set boundary conditions for this phase flow. Even in the simplest model, the dynamics of asteroids is non-trivial. An asteroid can start from tumbling and return back to tumbling without being disrupted, which contrasts the one-dimensional YORP model by \cite{pravec08}, where such behavior was impossible. Even more complicated dynamics can occur if TYORP, thermal lag or type III/IV asteroids are considered.

To give a taste of such complications, we consider one of them, namely TYORP, leaving the rest for the future. If TYORP is sufficiently large, equilibria between TYORP and NYORP arise, some of them stable. Asteroids can be attracted to such equilibria and kicked away from the overall evolution. This result agrees with the preliminary findings from \cite{golubov12} and \cite{golubov_lipatova}, and supplements the equilibria expected in more physically complicated models, such as binary asteroids \citep{golubov16equilibrium,golubov18}, tumbling asteroids \citep{breiter15} or asteroids with the thermal lag \citep{scheeres08}.

Although probability for an asteroid to reach an equilibrium in its $\varepsilon$-$\omega$ diagram can be relatively low, the asteroid can eventually reach it after undergoing several YORP cycles and enduring several alterations of its shape. This process looks like a ``natural selection'' of the asteroid shapes, in which only the ones allowing for stable equilibria survive, while the others are altered by centrifugal forces. Such YORP equilibria can be as important for distributing asteroids over rotation states, as the general properties of the phase flow and its boundary conditions. This leads us to a testable prediction that a significant fraction of asteroids would have YORP acceleration close to zero, and pushes us to pay more attention to negative detections of the YORP acceleration for asteroids.

%END Discussion

\section*{Acknowledgements}
O.G. acknowledges the help of Uliana Pyrohova, with whom he extensively discussed the program for computing the YORP coefficients of asteroids, and who also produced the preliminary version of Figure \ref{fig:yorp_obs}. The idea of this article and its first analytic results originated in discussions between O.G. and Veronika Lipatova. The first version of the program for computing evolution of the asteroid was co-written by Veronika Lipatova and O.G. \citep{golubov_lipatova}.

%BEGIN Appendix
\appendix

\section{Analytical theory for the YORP coefficients}
\label{app:taylor}

Neglecting the thermal inertia of the surface in the results of \citep{golubov16nyorp,golubov18erratum}, one gets the following expressions for the YORP torques acting on a convex asteroid:
\begin{align}
T_z = \frac{1}{R^3} \oint\limits_S \mathrm{d}S \, r\, \sin{\Delta}\cos{\eta}\cos{\psi} \, p^\alpha_z\left(\psi, \varepsilon\right),
\label{T_z_int}
\end{align}
\begin{align}
\label{T_epsilon_int}
T_{\varepsilon} = &-\frac{1}{R^3} \oint\limits_S \mathrm{d}S \, r\,\sin{\Delta}\cos{\eta}\sin{\psi}\, p^\alpha_\mathrm{sin}\left(\psi, \varepsilon\right)
\end{align}
Angles $\psi$, $\eta$, $\Delta$ are defined by the orientation of a surface element on the asteroid, and are explained in Figure \ref{fig:geometry}. $p^\alpha_z$ and $p^\alpha_\mathrm{sin}$ are the dimensionless YORP pressures, defined as follows:
\begin{figure}
	\includegraphics[width=0.48\textwidth]{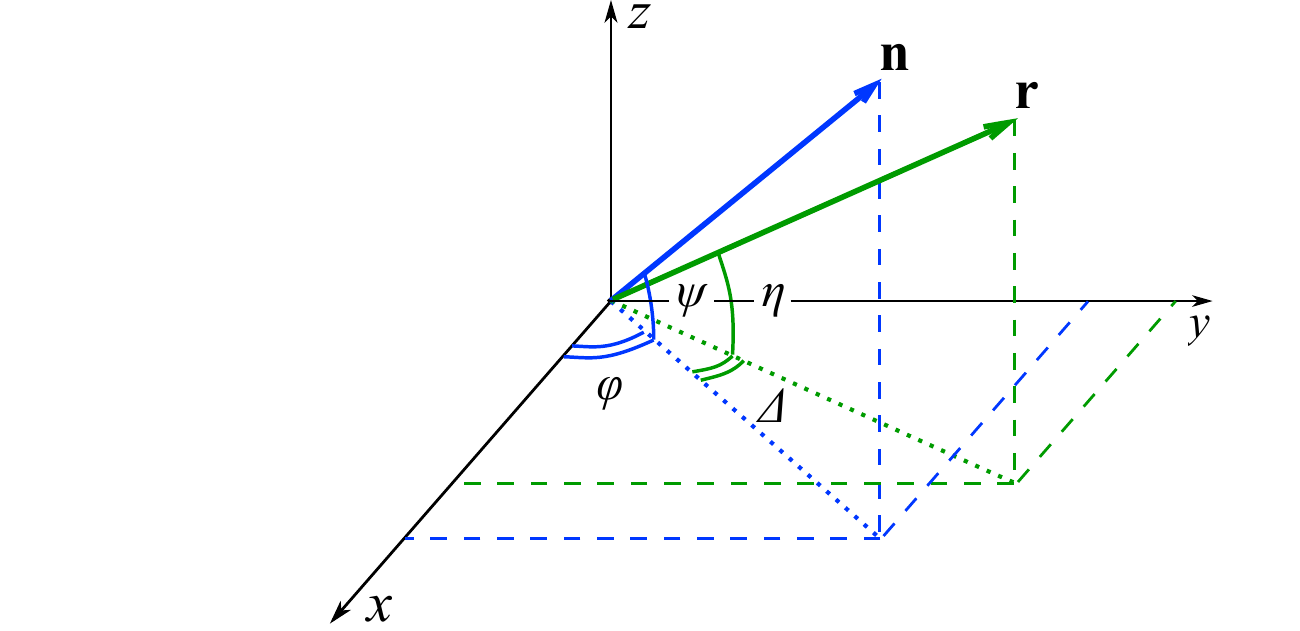}
	\caption{Orientation of the normal vector $\mathbf{n}$ and the radius vector $\mathbf{r}$ of the surface element with respect to the coordinate system.
		$\psi$ is the latitude of the surface element determined from its slope, $\psi$ is its latitude determined from the radius vector orientation.
		The angle $\phi$ between $Ox$ axis and the projection of $\mathbf{n}$ onto the equatorial plane $Oxy$ changes as the asteroid rotates, while the angle $\Delta$ between the projections of $\mathbf{r}$ and $\mathbf{n}$ remains constant.}
	\label{fig:geometry}
\end{figure}
\begin{align}
\label{p_a_z_simpl}
p^\alpha_z\left(\psi, \varepsilon\right) = &\frac{2}{3\pi^2} \int\limits^{\pi/2}_{-\pi/2} \mathrm{d}\phi \times \\
\nonumber & \times \sqrt{1 - \left(\sin{\phi}\cos{\psi}\sin{\varepsilon} - \sin{\psi}\cos{\varepsilon}\right)^2}\ ,
\end{align}
\begin{align}
\label{p_a_sin_simpl}
p^\alpha_\mathrm{sin}\left(\psi, \varepsilon\right) = &\frac{2}{3\pi^2} \int\limits^{\pi/2}_{-\pi/2} \mathrm{d}\phi \sin{\phi} \times \\
\nonumber & \times \sqrt{1 - \left(\sin{\phi}\cos{\psi}\sin{\varepsilon} - \sin{\psi}\cos{\varepsilon}\right)^2}\ .
\end{align}

Assuming that the square under the square root in Eqs. (\ref{p_a_z_simpl}) and (\ref{p_a_sin_simpl}) is in general much smaller than unity, we can decompose the square root into a Taylor series, keeping only the zeroth and the first order terms, and perform the integration analytically.
Thus we get the approximate equations
\begin{eqnarray}
\label{p_a_z_linear}
p^\alpha_z\left(\psi, \varepsilon\right) \approx \frac{2}{3\pi} \left( 1-\frac{1}{2}\sin^2\psi \cos^2\varepsilon-\frac{1}{4}\cos^2\psi \sin^2\varepsilon\right),
\end{eqnarray}
\begin{eqnarray}
\label{p_a_sin_linear}
p^\alpha_\mathrm{sin}\left(\psi, \varepsilon\right) \approx \frac{1}{12\pi} \sin{2\psi}\sin{2\varepsilon}\, .
\end{eqnarray}
Any summand independent of $\psi$ can be subtracted from $p^\alpha_z$ or $p^\alpha_\mathrm{sin}$, as it corresponds to an isotropic Pascal pressure upon the asteroid, which does not exert any torque. Thus we can express $\sin^2\psi$ through $\cos^2\psi$, express $\sin^2\varepsilon$ and $\cos^2\varepsilon$ in terms of $\cos 2\varepsilon$, neglect all the constants independent of $\psi$, which emerge after this substitution, and get an alternative expression for $p^\alpha_z$:
\begin{eqnarray}
\label{p_a_z_linear2}
p^\alpha_z\left(\psi, \varepsilon\right) \approx -\frac{1}{4\pi} \sin^2\psi\left(\cos2\varepsilon+\frac{1}{3}\right).
\end{eqnarray}
From Eqn. (\ref{p_a_z_linear2}) we see, that if $\varepsilon=\frac{1}{2}\arccos\left(-\frac{1}{3}\right)=54.7^\circ$, then $p^\alpha_z$ turns into a constant, and the integral in Eqn. (\ref{T_z_int}) vanishes -- thus confirming the well-known fact, that the axial component of YORP vanishes for obliquities about $\varepsilon=55^\circ$ \cite{rubincam00}.

Substituting Eqs. (\ref{p_a_z_linear2}) and (\ref{p_a_sin_linear}) into Eqs. (\ref{T_z_int}) and (\ref{T_epsilon_int}), we arrive at
\begin{align}
T_z &= -\frac{\Phi R^3}{c}\frac{1}{4\pi}\left(\cos2\varepsilon+\frac{1}{3}\right) \nonumber\\
&\times\frac{1}{R^3}\oint\limits_S \sin^2{\psi}\cos{\psi}\cos{\eta}\sin{\Delta}\,r\,\mathrm{d}S\,.
\label{T_z_int_correlation}
\end{align}
\begin{align}
T_{\varepsilon} &= -\frac{\Phi R^3}{c}\frac{1}{6\pi}\sin{2\varepsilon} \nonumber\\
&\times\frac{1}{R^3}\oint\limits_S \sin^2{\psi}\cos{\psi}\cos{\eta}\sin{\Delta}\,r\,\mathrm{d}S\,.
\label{T_epsilon_int_correlation}
\end{align}

We introduce the notation
\begin{equation}
C_\varepsilon = -\frac{1}{6\pi}\frac{1}{R^3}\oint\limits_S \sin^2{\psi}\cos{\psi}\cos{\eta}\sin{\Delta}\,r\,\mathrm{d}S\,.
\label{T_epsilon_int_correlation}
\end{equation}
Then Eqs. (\ref{T_z_int_correlation}) and (\ref{T_epsilon_int_correlation}) turn into
\begin{equation}
T_z = \frac{3}{2}\frac{\Phi R^3}{c}C_\varepsilon\left(\cos2\varepsilon+\frac{1}{3}\right)\,,
\end{equation}
\begin{equation}
T_{\varepsilon} = \frac{\Phi R^3}{c}C_\varepsilon\sin{2\varepsilon}\,.
\end{equation}
This coincides with Eqs. (\ref{T_z_cos}) and (\ref{T_eps_sin}), assuming the coefficients $\alpha=C_z/C_\varepsilon=\frac{3}{2}$ and $\beta=\frac{1}{3}$.
These two coefficients are marked with a black square in Figure \ref{fig:alpha-beta}. 
Although $\beta=\frac{1}{3}$ agrees with the data points for real asteroids, $\alpha=\frac{2}{3}$ is still about 10\% too small and lies outside the flock of data points.
To correct this discrepancy, the theory should be taken to higher orders.

\section{Fitted model for the YORP coefficients}
\label{app:lsf}
The Taylor decomposition we used to transform Eqs. (\ref{p_a_z_simpl}) and (\ref{p_a_sin_simpl}) into Eqs. (\ref{p_a_z_linear}) and (\ref{p_a_sin_linear}) is precise only for some particular values of the angles $\phi$, $\psi$ and $\varepsilon$. Presumably, if we fit the exact formulas with approximations in the form of Eqs. (\ref{p_a_sin_linear}) and (\ref{p_a_z_linear2}), but with free fitting coefficients, we can achieve a better precision for the calculated YORP effect. Thus we choose the following form of $p^\alpha_z$ and $p^\alpha_\mathrm{sin}$:
\begin{equation}
\label{p_a_z_lsm}
p^\alpha_z\left(\psi, \varepsilon\right) \approx b \sin^2\psi\cos2\varepsilon+c\sin^2\psi+F(\varepsilon)\,,
\end{equation}
\begin{equation}
\label{p_a_sin_lsm}
p^\alpha_\mathrm{sin}\left(\psi, \varepsilon\right) \approx a \sin{2\psi}\sin{2\varepsilon}\, .
\end{equation}
Here $a$, $b$ and $c$ are three fitting parameters, and $F(\varepsilon)$ is a fitting function.
We do not add a free fitting function in Eqn(\ref{p_a_sin_lsm}), as the mean of $p^\alpha_\mathrm{sin}$ after averaging over $\psi$ is 0, so that this free function would vanish for any reasonable kind of fitting.

For fitting we use the least squares method, over the range $0<\psi<\frac{\pi}{2}$, $0<\varepsilon<\frac{\pi}{2}$.
Given the symmetry of the fitted functions (Eqs. (\ref{p_a_z_simpl}) and (\ref{p_a_sin_simpl})) and the fitting functions (Eqs. (\ref{p_a_z_lsm}) and (\ref{p_a_sin_lsm})), it is equivalent to fitting over the whole range of variables $-\frac{\pi}{2}<\psi<\frac{\pi}{2}$, $-\frac{\pi}{2}<\varepsilon<\frac{\pi}{2}$.

Not all the points are equivalent for this fitting. The biggest contribution to $T_z$ of the entire asteroid is provided by $\psi\approx 0$, while $T_\varepsilon$ is the most strongly influenced by the points with $\psi\approx \frac{\pi}{4}$. Therefore, these points should get a higher weight in the averaging.

We estimate the proper averaging weights from Eqs. (\ref{T_z_int}) and (\ref{T_epsilon_int}) by approximately taking $\eta\approx\psi$, $\mathrm{d}S\propto\cos\psi$. Moreover, if $\delta$ is the random scatter of orientations of facets in three dimensions, then we can estimate $\Delta\approx\delta/\cos\psi$. Substituting it into Eqs. (\ref{T_z_int}) and (\ref{T_epsilon_int}), we see that for each particular value of $\psi$, $p^\alpha_z$ enters the integral Eqs. (\ref{T_z_int}) with the factor approximately proportional to $\cos^2\psi$, while $p^\alpha_\mathrm{sin}$ enters the integral Eqs. (\ref{T_epsilon_int}) with the factor approximately proportional to $\cos\psi\sin\psi$. These are the factors we take for the inverse errors in the least squares fitting.

In such a way, after numerically computing the integrals in Eqs. (\ref{p_a_z_simpl}) and (\ref{p_a_sin_simpl}), we get the best fit $a=0.0395$, $b=-0.1093$, $c=-0.0355$. They correspond to $\alpha=0.722$, $\beta=0.325$, which are plotted in Figure \ref{fig:alpha-beta} with a black circle. This estimate is in much better agreement with the observed data. The value of $\beta=0.325$ corresponds to $T_z$ turning to zero at $\varepsilon=54.5^\circ$.

\section{Handling of asteroid shape models}
\label{app:models}

For the following analysis we use asteroid shapes from different sources: photometric observations, radar measurements and \textit{in situ} observations.

Photometric shape models were collected from the DAMIT database via complete data flush on December 12, 2017 \cite{damit}.
They were derived by the supporters of the database from photometric data by the lightcurve inversion method, and in some cases later improved using adaptive optics images, infrared observations, or occultation data.
The database includes 1706 models of 943 asteroids. The majority of the models, if not all, are convex.
In the further analysis, we treat all the 1706 models separately, irrespective to whether they represent different shape solutions for the same asteroid.

The radar shape models were taken from the JPL Asteroid Radar Research website \cite{radar}.
The database included 26 shape models, which are in general non-convex.

Models from \textit{in situ} observations of asteroids 433 Eros \cite{gaskel_eros} and 25143 Itokawa \cite{gaskel_itokawa} were also included into our analysis.
For each of these asteroids we used four models of different resolution, all treated separately.

\begin{figure}
	\centering
	\includegraphics[width=0.375\textwidth]{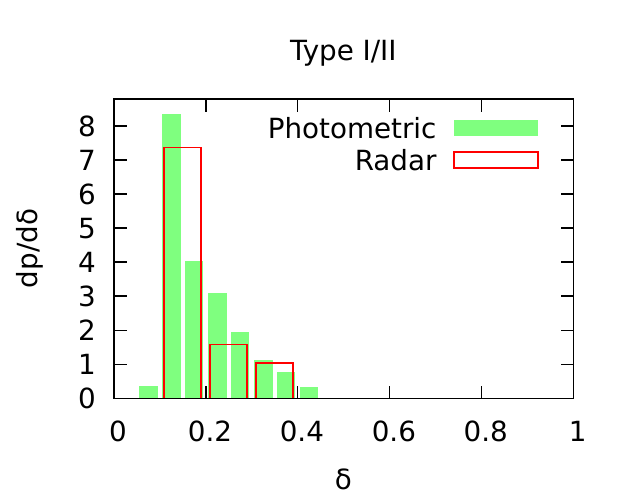}
	\includegraphics[width=0.315\textwidth]{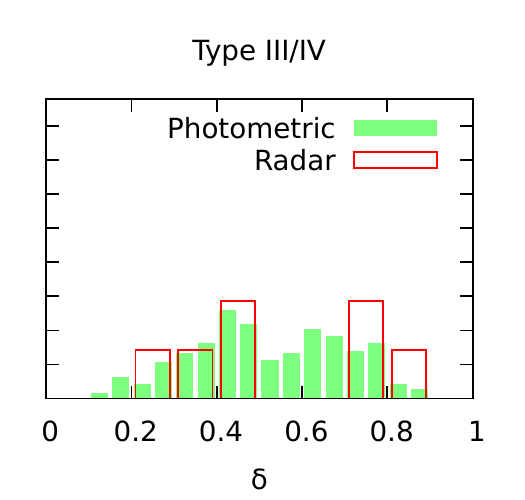}
	\caption{Histograms showing the distribution of the asteroid shape models over $\delta$. The left-hand panel shows type I/II shape models \citep{vokrouhlicky02}, the right-hand panel shows type III/IV shape models. Photometric shape models are shown in green, radar shape models in red.}
	\label{fig:delta}
\end{figure}

\begin{figure*}
	\centering
	\includegraphics[width=0.48\textwidth]{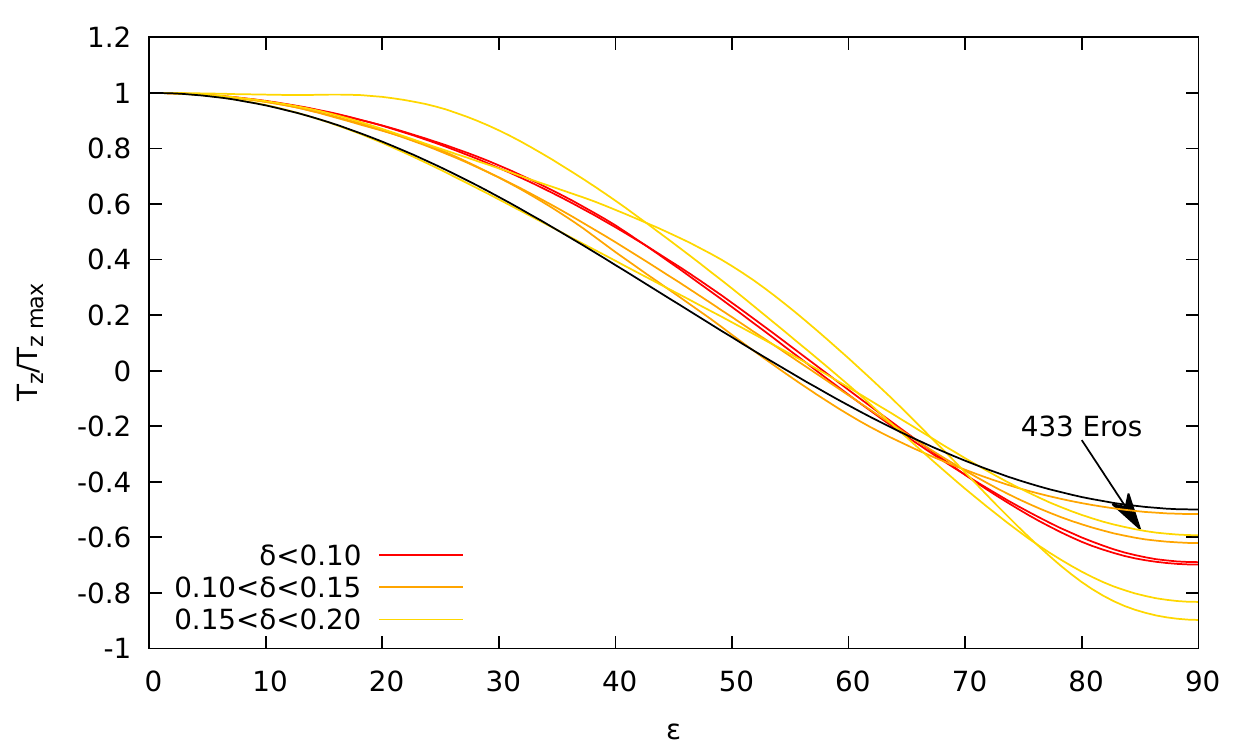}
	\includegraphics[width=0.48\textwidth]{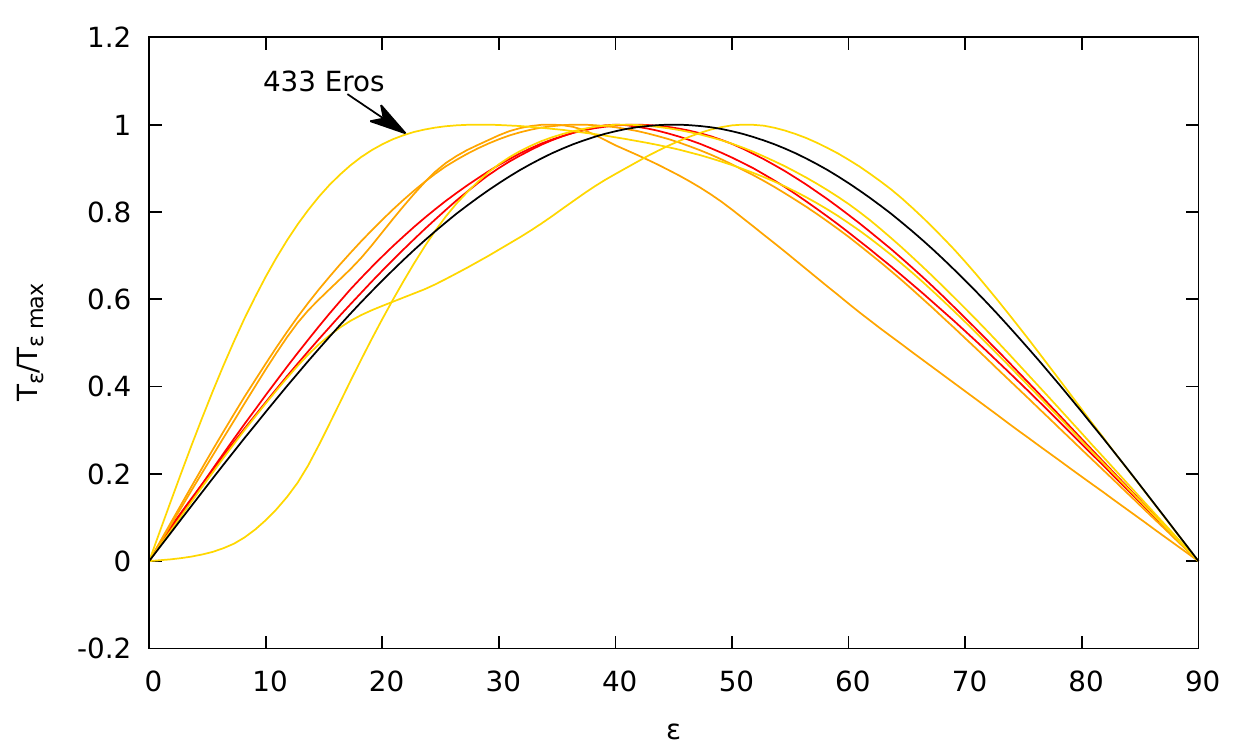}
	\includegraphics[width=0.48\textwidth]{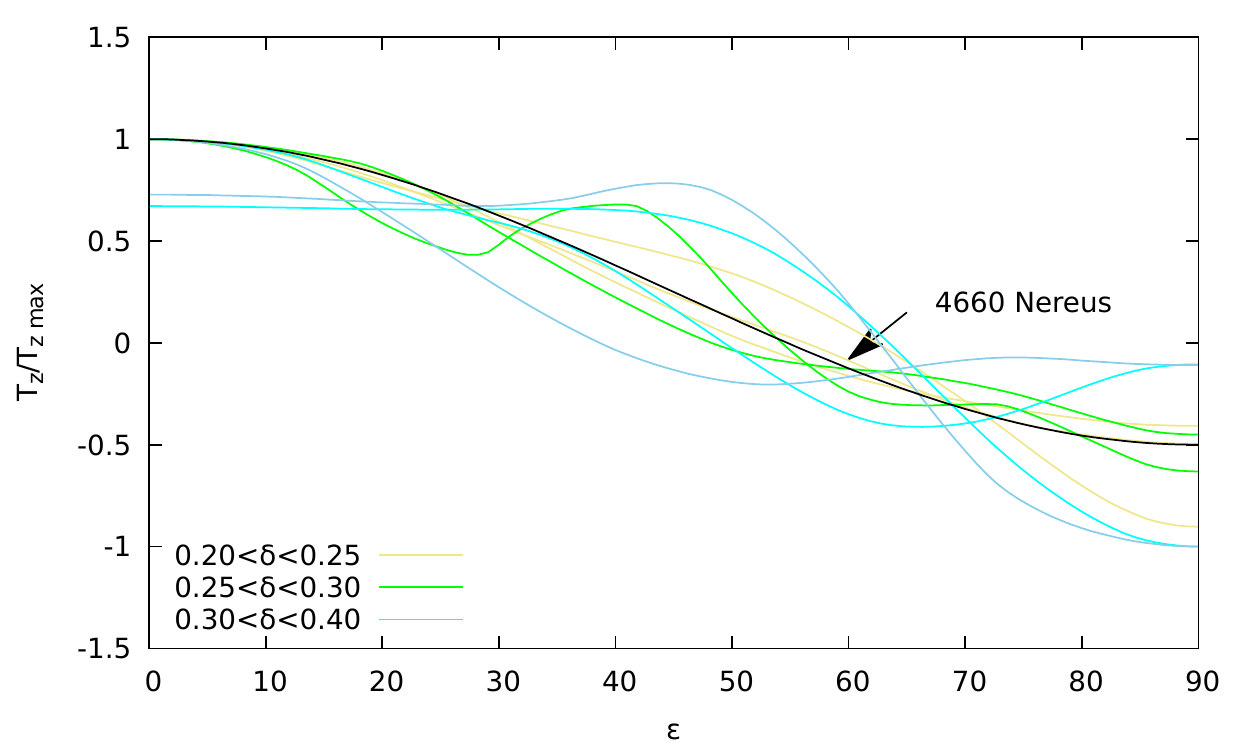}
	\includegraphics[width=0.48\textwidth]{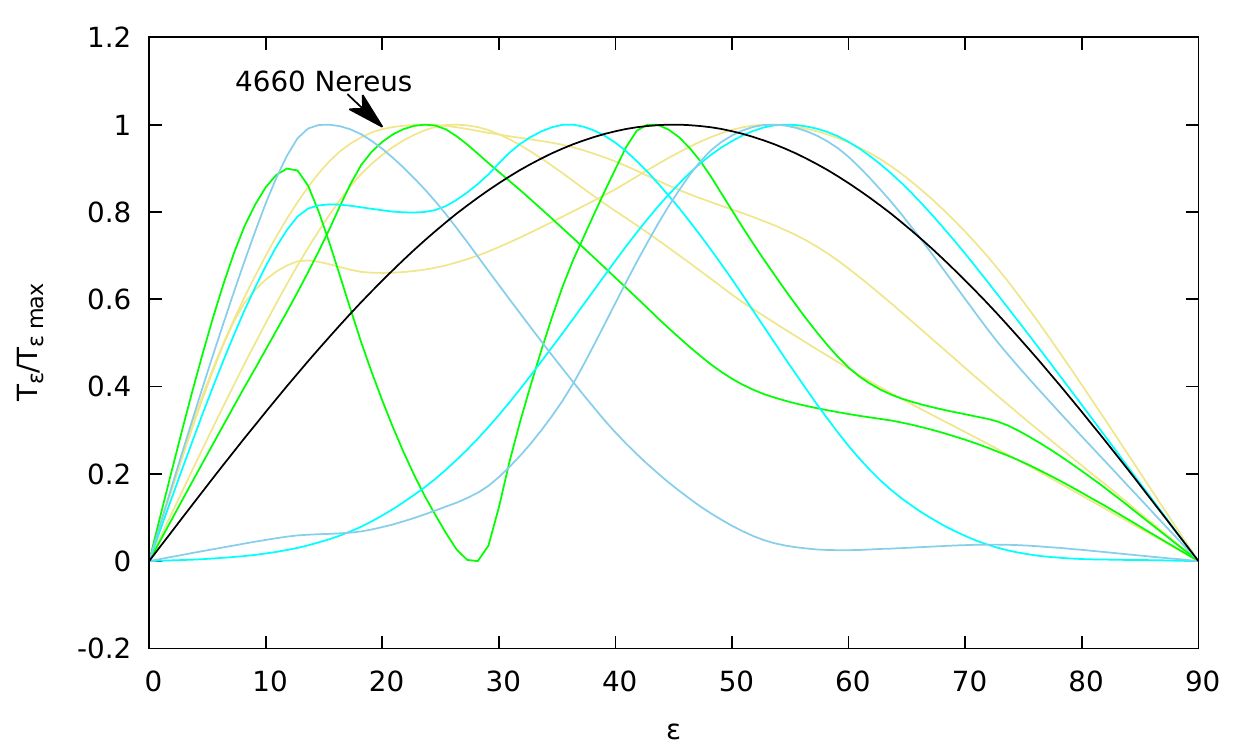}
	\includegraphics[width=0.48\textwidth]{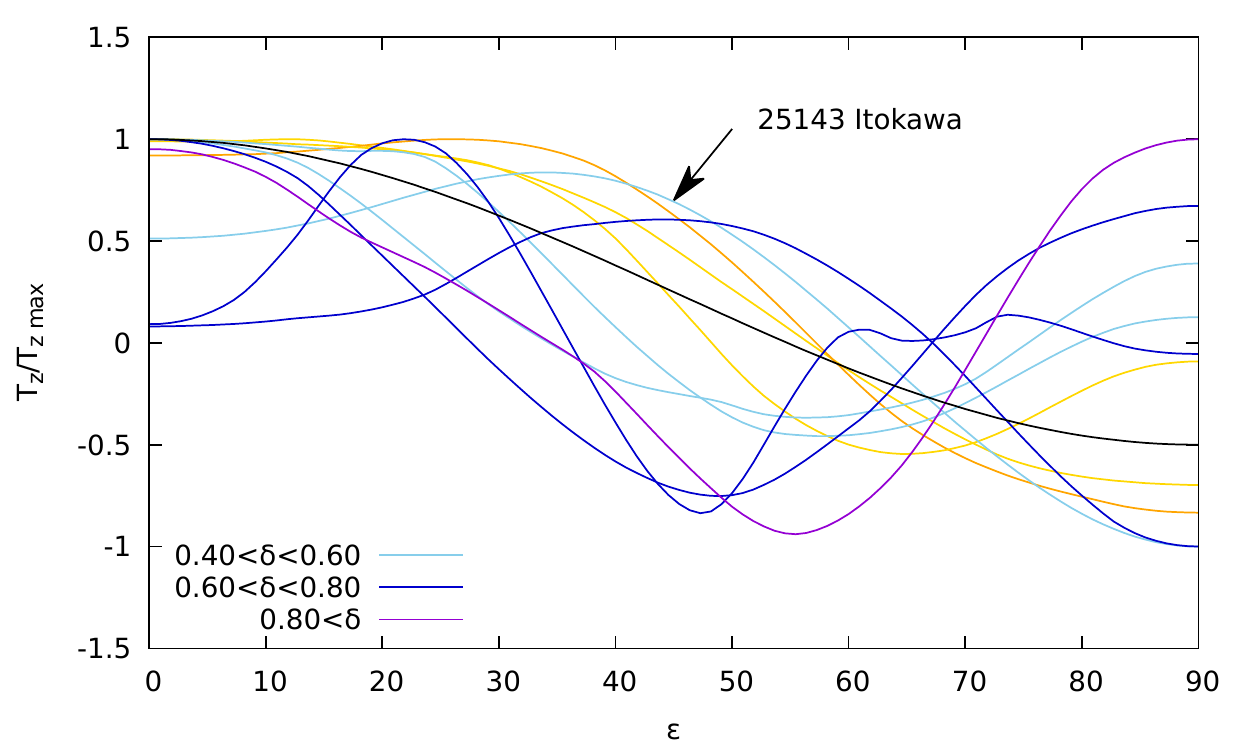}
	\includegraphics[width=0.48\textwidth]{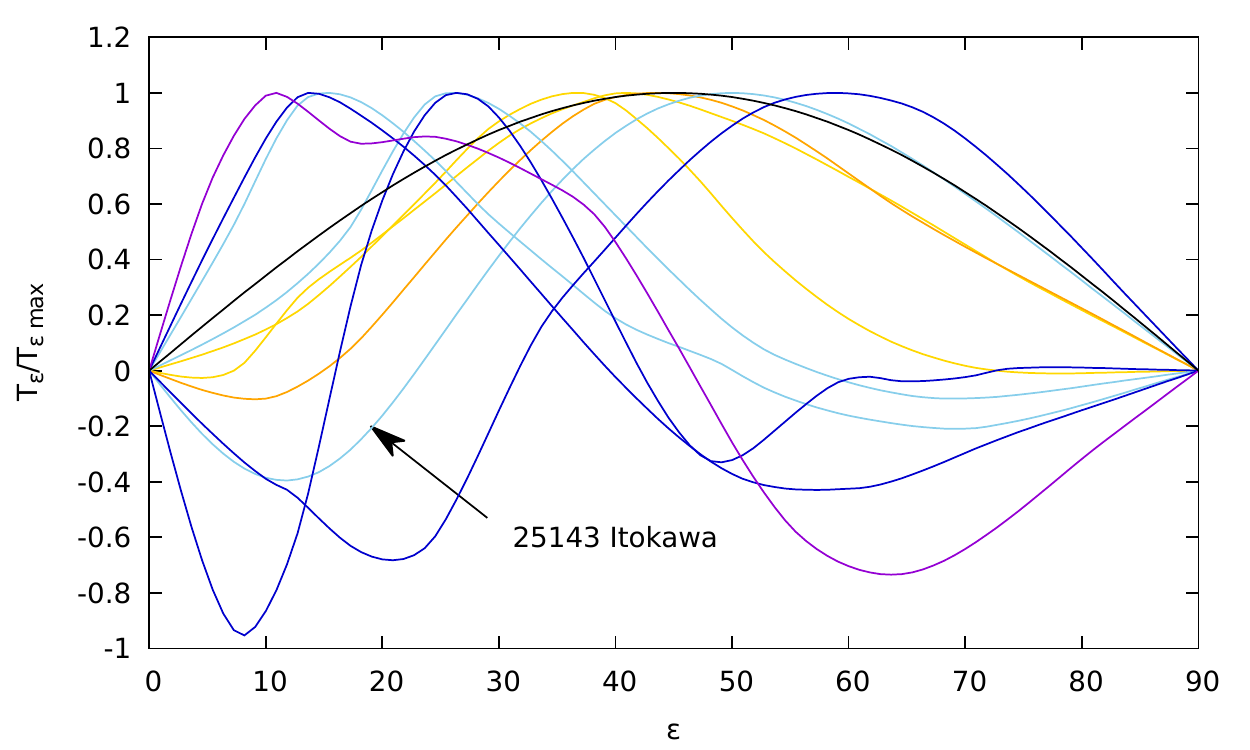}
	\caption{Axial (left column) and obliquity (right column) components of YORP as a function of obliquity $\varepsilon$ for different asteroids. The top panels show type I/II asteroids with $\delta<0.2$, the middle panels show type I/II asteroids with $\delta>0.2$, the bottom panels -- type III/IV asteroids.}
	\label{fig:yorp_obs}
\end{figure*}

\begin{figure}
	\centering
	\includegraphics[width=0.48\textwidth]{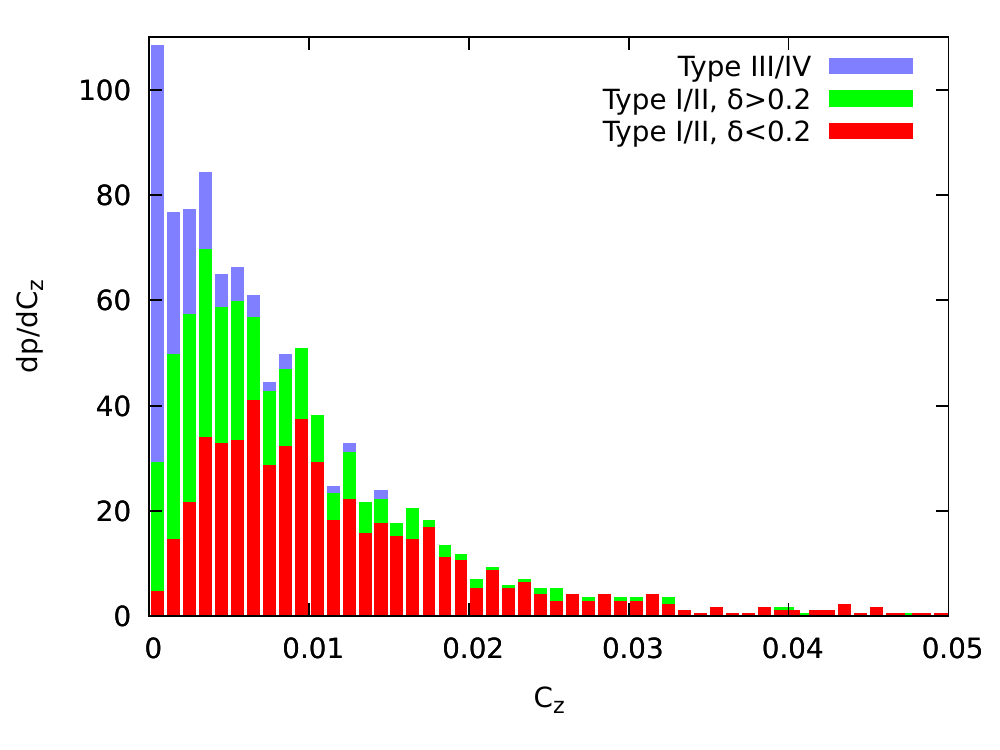}
	\caption{Distribution of asteroids over the YORP coefficients. The total number of asteroids is split according to the discrepancy between the YORP with the trigonometric fit, the three different bins being shown in different colors.}
	\label{fig:distribution}
\end{figure}

We use the shape models exactly as they are in the data base, without reduction to the center of mass or the principal axes, as well as without checking for triangulation errors.
For all the available shape models, we compute the YORP effect as a function of obliquity, using Eqs. (\ref{T_z_int}) and (\ref{T_epsilon_int}).
We fit these curves by Eqs. (\ref{T_z_cos}) and (\ref{T_eps_sin}), using the least squares method.
The fineness of fit ranging from 0 to 1 is color-coded, so that the red curves are the best approximated by Eqs. (\ref{T_z_cos}) and (\ref{T_eps_sin}),
while purple curves go astray from the fit.

The parameters of the least squares fit are used to evaluate the parameters $C_z$, $C_\varepsilon$ and $\beta$ for Eqs. (\ref{T_z_cos}) and (\ref{T_eps_sin}). The residual of the fit is described as follows. First, we introduce $\delta_z$ and $\delta_\varepsilon$, the mean squared discrepancies between the modeled $T_z(\varepsilon)$ and $T_\varepsilon(\varepsilon)$ and their fits Eqs. (\ref{T_z_cos}) and (\ref{T_eps_sin}), normalized over the maximal absolute value of the model,
\begin{align}
\delta_z^2=\frac{1}{T_{z\,\mathrm{max}}^2}\frac{1}{\pi} \int\limits^{\pi}_{0}\left(T_z(\varepsilon)-C_z\cos{\varepsilon}-\beta\right)^2\,\mathrm{d}\varepsilon \nonumber \\
\delta_\varepsilon^2=\frac{1}{T_{\varepsilon\,\mathrm{max}}^2}\frac{1}{\pi} \int\limits^{\pi}_{0}\left(T_\varepsilon(\varepsilon)-C_\varepsilon\sin{\varepsilon}\right)^2\,\mathrm{d}\varepsilon
\end{align}
Then, the quality of the entire fit is described by $\delta=\sqrt{\delta_z^2+\delta_\varepsilon^2}$. The bigger is $\delta$, the worse is the fit. The distribution of asteroids over types and the value of $\delta$ is illustrated by Figure \ref{fig:delta} and Table \ref{table1}.

\begin{table}
\caption{Distribution of asteroids over types and fineness of the trigonometric fit}
\begin{tabular}{l|ccc}
Dataset & \multicolumn{2}{c}{Type I/II} & Type III/IV\\
& $\delta<0.2$ & $\delta>0.2$ \\
\hline
DAMIT & 53\% & 30\% & 17\% \\
Radar & 54\% & 19\% & 27\%
\end{tabular}
\label{table1}
\end{table}

Some sample plots of YORP as a function of obliquity are shown in Figure \ref{fig:yorp_obs}, the axial and the obliquity components in different panels. Photometric, radar and \textit{in situ} shapes are plotted with different line types. Marked are several individual asteroids, which are important for our discussion. Different ranges of $\delta$ are shown with different colors. Separately plotted are type I/II asteroids with small $\delta$, type I/II asteroids with large $\delta$, and type III/IV asteroids. Naturally, the quality of the fit is good for the top panels, worse for the middle panel, and the worst for the bottom panel. The plots are normalized in such a manner that the largest absolute value for each line is 1. The standard theoretical curves corresponding to $\beta=1/3$ are shown in black lines. They do not necessarily provide the best fit to the modeled $T_z(\varepsilon)$ and $T_\varepsilon(\varepsilon)$, firstly, because $\beta$ can be different, and, secondly, because the plots are normalized so that their maxima are the same, not their fits.

$C_\varepsilon/C_z$ and $\beta$ are approximately constant, so that to the first approximation the YORP effect of an asteroid is fully characterized by its $C_z$. In Figure \ref{fig:distribution} we show the distribution of asteroids in the DAMIT sample over $C_z$. We see, that small absolute values $|C_z|$ are more probable. Positive and negative signs of $C_z$ are equally probable. Type III/IV asteroids have smaller $|C_z|$ than type I/II asteroids. Asteroids with larger $\delta$ have smaller $|C_z|$. Overall, if an asteroid has a large absolute value $|C_z|$ and thus a large YORP effect, its YORP is well described by Eqs. (\ref{T_z_int}) and (\ref{T_epsilon_int}).

\section{Existence of equilibria with tangential YORP}
\label{app:equilibria}
In contrast to the obliquity component, the axial component of the YORP effect 
is independent of the rotation rate whenever the surface of the asteroid is locally flat,
so that one-dimensional heat conductivity model can be used for soil \citep{bbc10,golubov16nyorp}.
On the other hand, when non-flatness of the surface is substantial,
a new component of YORP arises, the tangential YORP, or TYORP \citep{golubov12,golubov14,sevecek15}.
TYORP operates only in some range of spins, being very small at large and small rotation rates.
Its contribution is always directed towards the increase of the absolute value of the rotation rate.

We use several different articles to assemble an analytic expression for the tangential YORP.
Firstly, \cite{golubov14} derive and \cite{sevecek16} confirm, that TYORP of an asteroid with zero obliquity can be expressed as $T_z\approx 9p\frac{\Phi R^3}{c}$, where $p$ is the dimensionless pressure at the equator.
Secondly, \cite{sevecek16} find that TYORP as a function of obliquity is approximately proportional to the factor $1+\cos^2\varepsilon$.
Thirdly, \cite{golubov17} derives an approximate analytic expression for $p$, namely $n_0 \mu\exp{\left(-\frac{\left(\ln{\theta}-\ln{\theta_0}\right)^2}{\nu^2}\right)}$. The constant $n_0$ is proportional to the number of boulders on the surface, while the three other constants are $\mu=0.00644$, $\nu=1.518$, $\ln{\theta_0}=0.580$ (assuming boulder size distribution with the power index $\gamma=-3$).
Assembling these three results together, we get the following expression for TYORP:
\begin{eqnarray}
T_{z\,\mathrm{TYORP}}=4.5\frac{\Phi R^3}{c} n_0 \mu\exp\left(-\frac{\left(\ln{\theta}-\ln{\theta_0}\right)^2}{\nu^2}\right)\times\nonumber\\
\times(1+\cos^2\varepsilon)\mathrm{sgn}(\omega)
\label{tyorp_torque}
\end{eqnarray}
The factor $\mathrm{sgn}(\omega)$ equals 1 if $\omega>0$ and equals $-1$ if $\omega<0$, so that TYORP always increases the absolute value of $\omega$. The thermal parameter $\theta$ entering this equation is defined as follows:
\begin{equation}
\theta = \frac{\left(C\rho\omega\kappa\right)^{1/2}}{\left(\varepsilon\sigma\right)^{1/4}\left(1-A\right)^{3/4}\Phi^{3/4}}.
\label{theta_definition}
\end{equation}
Here $A$ as the albedo, $\varepsilon$ is the thermal emissivity, $\sigma$ is the Stefan--Boltzmann constant,
$\kappa$ is the heat conductivity of the material constituting the asteroid surface, $\rho$ its density,
$C$ is its specific heat capacity, and $\Phi$ the solar irradiance at the asteroid's distance.

For the normal component of YORP we assume the generic behavior given by Eqn. (\ref{T_z_cos}) with the constant $\beta=0.33$, namely
\begin{equation}
\tau_{z\,\mathrm{NYORP}}=\frac{\Phi R^3}{c}C_\omega(\cos 2\varepsilon+\beta).
\label{tyorp_torque}
\end{equation}

An equilibrium rotation state of an asteroid is defined by the following set of equations:
\begin{align}
T_{\varepsilon}=0, \label{Teps=0}\\
T_{z\,\mathrm{NYORP}}+T_{z\,\mathrm{TYORP}}=0. \label{Tomega=0}
\end{align}
The first equation is satisfied for $\varepsilon=0^\circ$, $90^\circ$ and $180^\circ$. Let us consider the possible equilibrium values of $\varepsilon$ in turn.

Firstly, if $\varepsilon=0^\circ$ or $180^\circ$, then $T_{z\,\mathrm{NYORP}}=\frac{\Phi R^3}{c}\frac{1+\beta}{\alpha}C_\varepsilon$ is positive, therefore $T_{z\,\mathrm{TYORP}}$ must be negative for Eqn. (\ref{Tomega=0}) to be satisfied. Then Eqn. (\ref{tyorp_torque}) implies that $\omega$ is negative. The largest possible absolute value of TYORP allowed by Eqn. (\ref{tyorp_torque}) at $\varepsilon=0^\circ$ or $180^\circ$ is $9\frac{\Phi R^3}{c} n_0 \mu$. It must be larger than $T_{z\,\mathrm{TYORP}}$ for the equilibria to exist, or equivalently
\begin{equation}
C_\omega<\frac{9n_0\mu}{(1+\beta)}.
\end{equation}

Secondly, if $\varepsilon=90^\circ$, then $T_{z\,\mathrm{NYORP}}=-\frac{\Phi R^3}{c}C_\omega(1-\beta)$ is negative, therefore at equilibrium $T_{z\,\mathrm{TYORP}}$ and $\omega$ must be positive. The largest possible TYORP at $\varepsilon=90^\circ$ is $4.5\frac{\Phi R^3}{c} n_0 \mu$, and the condition for the equilibrium results into
\begin{equation}
C_\omega<\frac{9n_0\mu}{2(1-\beta)}.
\end{equation}

After substituting the numerical values of the coefficients, the two conditions for different signs of $C_\varepsilon$ unite into one equation,
\begin{equation}
C_\omega<0.04n_0.
\label{tyorp_equilibria_condition}
\end{equation}
The equilibria at $\varepsilon=0^\circ$, $90^\circ$ and $180^\circ$ appear simultaneously, because the absolute values of both TYORP and NYORP at $90^\circ$ are exactly 2 times smaller, than at $0^\circ$ and $180^\circ$. Naturally, Eqs. (\ref{Teps=0}) and (\ref{Tomega=0}) are just approximations, so that this degeneration holds also only approximately.

The value of $n_0$ for 25143 Itokawa is about 0.03 \citep{sevecek16}. Assuming that this value of $n_0$ is typical for asteroids, we get the condition $C_\omega<0.001$. It is about an order of magnitude less than the typical value of $C_\omega$ seen in Figure \ref{fig:distribution}.
Thus we expect only of the order of 10\% of asteroids to be capable of achieving this equilibrium between NYORP and TYORP.

\begin{figure}
	\centering
	\includegraphics[width=0.48\textwidth]{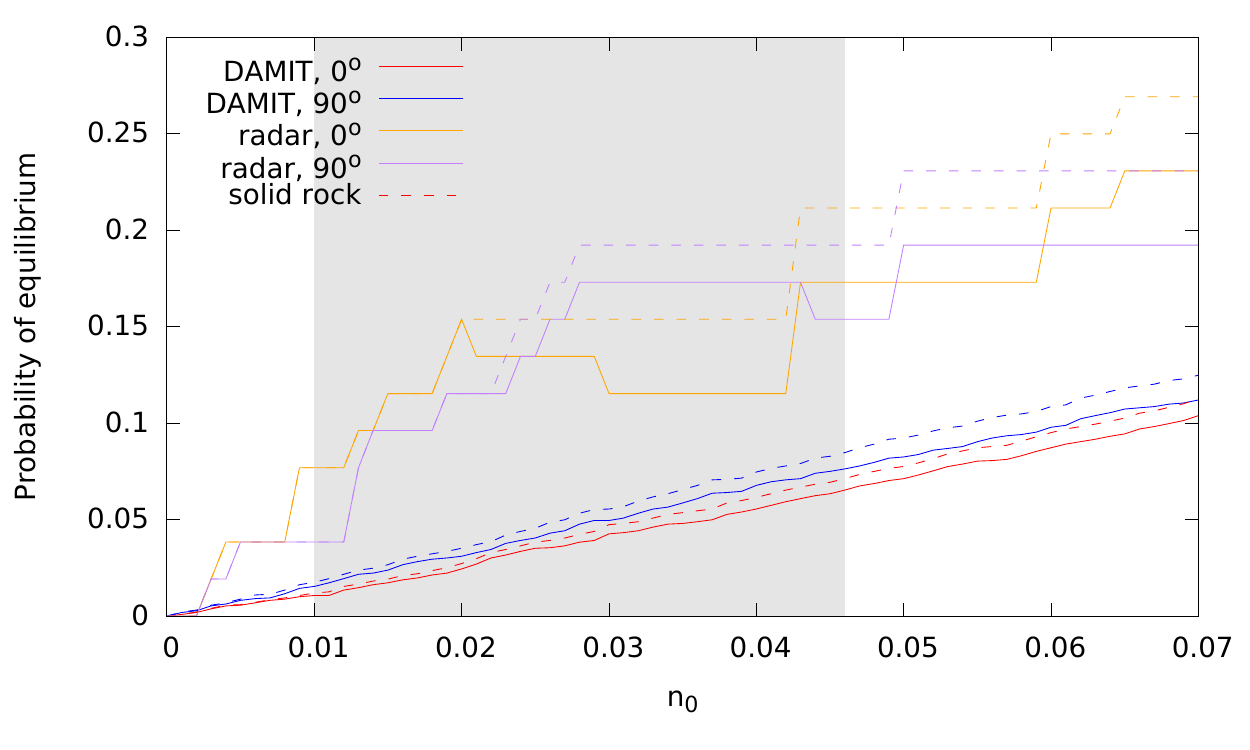}
	\caption{Probability of equilibria as a function of the surface density of boulders. Shaded area swows the uncertainty range for the surface density of boulders on the surface of 25143 Itokawa.}
	\label{fig:tyorp-probability}
\end{figure}

\begin{figure}
	\centering
	\includegraphics[width=0.48\textwidth]{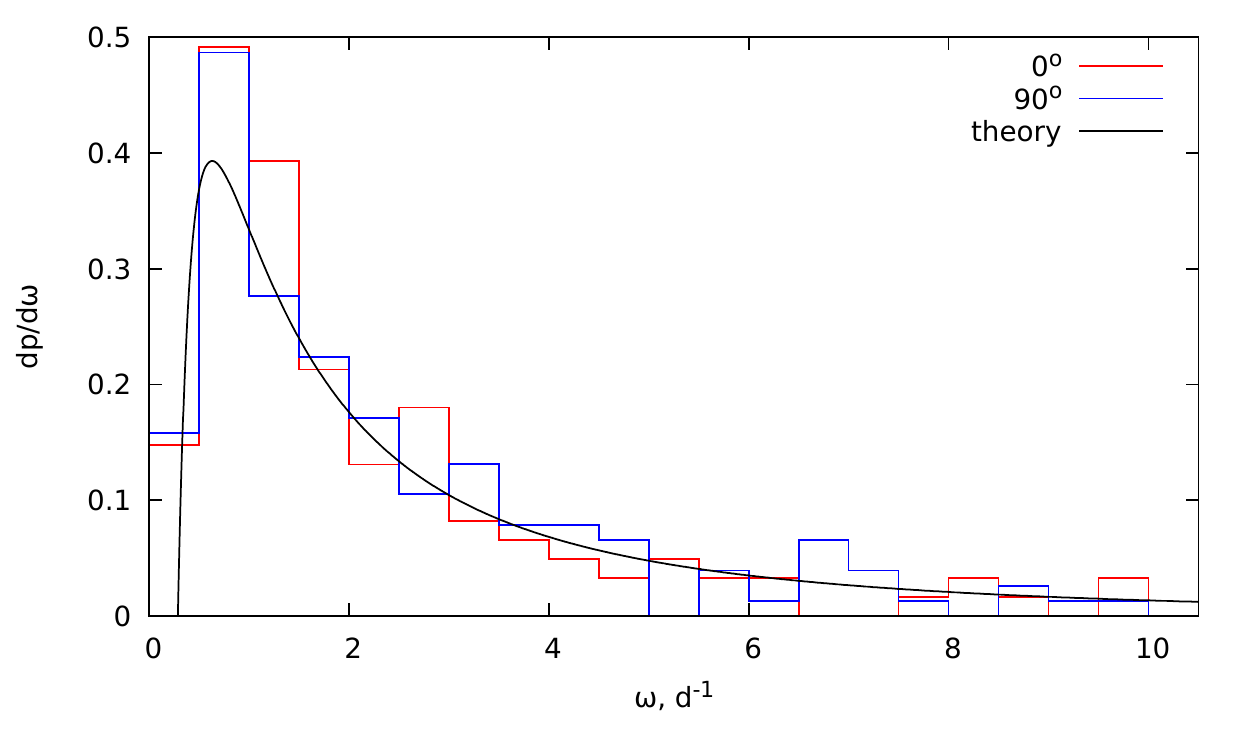}
	\includegraphics[width=0.48\textwidth]{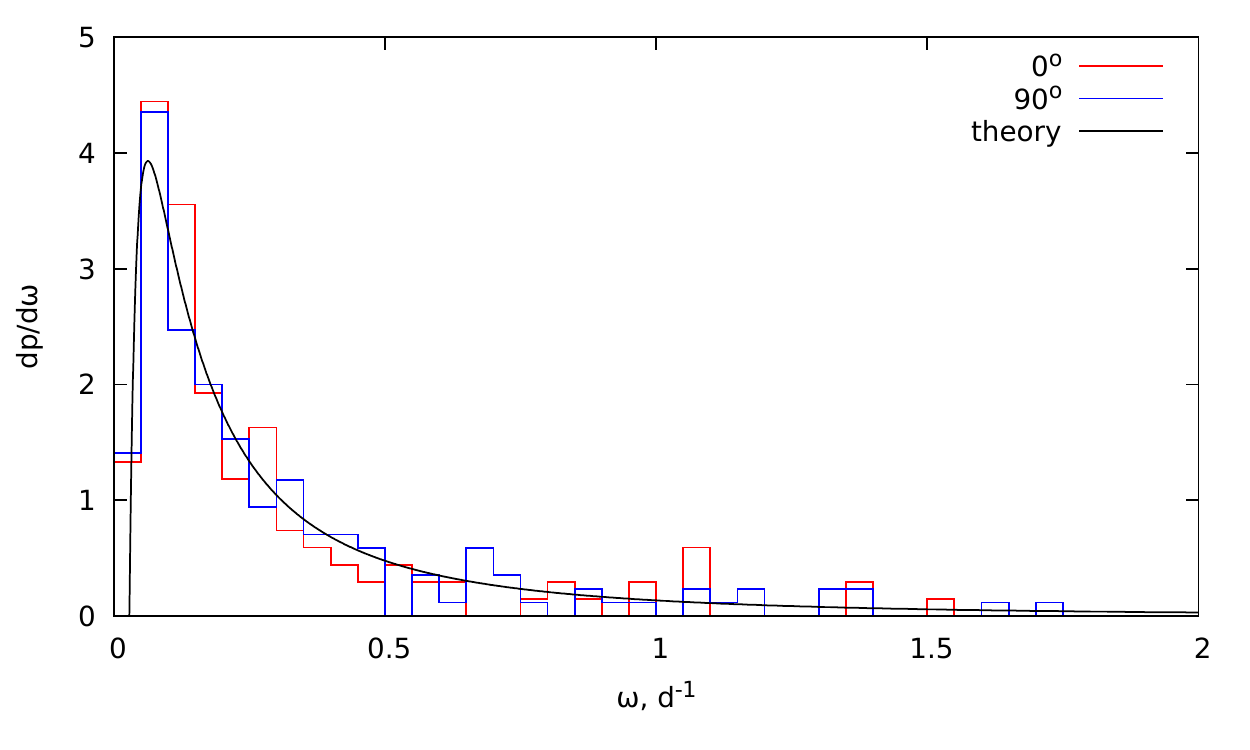}
	\caption{Distribution of asteroids over equilibrium rotation rates. Surface density $n_0=0.028$ is assumed. The left panel corresponds to cracked rock, the right panel to solid rock.}
	\label{fig:tyorp-distribution}
\end{figure}

\end{document}